\documentclass[twocolumn,showpacs,preprintnumbers,amsmath,amssymb,aps,pre]{revtex4}
\usepackage[dvips]{graphicx}% Include figure files
\usepackage{dcolumn}% Align table columns on decimal point
\usepackage{bm}% bold math
\newcommand{\p}{\partial}
\newcommand{\rd}{\mathrm{d}}
\newcommand{\e}{\mathrm{e}}

\newcommand{\px}{{\bm{x}}}
\newcommand{\ps}{{\bm{s}}}
\newcommand{\pF}{{\bm{F}}}
\renewcommand{\(}{\left(}
\renewcommand{\)}{\right)}
\renewcommand{\[}{\left[}
\renewcommand{\]}{\right]}

\newcommand{\prt}[2]{\frac{\partial #1}{\partial #2}}
\newcommand{\prts}[3]{\frac{\partial^{#3} #1}{\partial {#2}^{#3}}}

\newcommand{\expct}[1]{\langle #1 \rangle}
\newcommand{\Expct}[1]{\left\langle #1 \right\rangle}

\newcommand{\Cumulant}[2]{\left\langle #1,#2 \right\rangle}
\renewcommand{\eqref}[1]{Eq.~(\ref{#1})}
\newcommand{\eqsref}[1]{Eqs.~(\ref{#1})}
\newcommand{\pref}[1]{(\ref{#1})}
\DeclareMathOperator{\sgn}{sgn}
\DeclareMathOperator{\Arcsinh}{Arcsinh}  %Arcsin
\allowdisplaybreaks[1]

\begin{document}

\preprint{APS/123-QED}

\title{Role of unstable periodic orbits in phase transitions of coupled map lattices}% Force line breaks with \\

\author{Kazumasa Takeuchi}
 \email{kazumasa@daisy.phys.s.u-tokyo.ac.jp}
\author{Masaki Sano}%
\affiliation{%
Department of Physics, The University of Tokyo, 7-3-1
  Hongo, Bunkyo-ku, Tokyo, 113-0033, Japan
}%

\date{\today}% It is always \today, today,
             %  but any date may be explicitly specified

\begin{abstract}
The thermodynamic formalism for dynamical systems with many degrees of freedom
 is extended to deal with time averages and fluctuations
 of some macroscopic quantity along typical orbits,
 and applied to coupled map lattices exhibiting phase transitions.
Thereby, it turns out that a seed of phase transition
 is embedded as an anomalous distribution of unstable periodic orbits,
 which appears as a so-called q-phase transition
 in the spatio-temporal configuration space.
This intimate relation between phase transitions and q-phase transitions
 leads to one natural way of defining transitions and their order
 in extended chaotic systems.
Furthermore, a basis is obtained on which we can treat
 locally introduced control parameters as macroscopic ``temperature''
 in some cases involved with phase transitions.
\end{abstract}

% PACS, the Physics and Astronomy
                             % Classification Scheme.
%\keywords{Suggested keywords}%Use showkeys class option if keyword
                              %display desired
\maketitle

\section{Introduction}  \label{sec:1}

For the past few decades
 chaotic dynamical systems with a few degrees of freedom (DOFs)
 have been investigated theoretically, numerically, and experimentally
 with enthusiasm, which has brought various insights about them.
Since one cannot follow individual trajectories in chaotic systems
 by any means,
 one of the subjects attracting interest is evaluation of dynamical averages,
 namely asymptotic time averages and fluctuations of some observables
 along typical orbits.
The thermodynamic formalism \cite{Ruelle-1,Beck-1},
 which is frequently used for multifractal analysis
 \cite{Beck-1,Halsey_etal-1},
 is exploited for this purpose \cite{Fujisaka_Inoue-1}
 and the concept of dynamical averaging has been remarkably developed
 by means of unstable periodic orbit expansion, trace formulae,
 and dynamical zeta function,
 which reveal the role of unstable periodic orbits (UPOs)
 as a skeleton of chaos \cite{Cvitanovic_etal-1}.
On the other hand, the thermodynamic formalism is sometimes discussed
 in the context of phase transitions, called q-phase transitions.
This is not a transition dealt with in statistical mechanics
 which involves large fluctuations of thermodynamic quantities
 and occurs only in the thermodynamic limit,
 but a transition with large dynamical fluctuations of observables
 which occurs in the long-time limit.
It has been shown that the large fluctuations reflect the dynamics
 and q-phase transitions indicate a singular local structure
 of the chaotic attractor, such as homoclinic tangencies
 of stable and unstable manifolds and band crises \cite{Hata_etal-1}.

Despite the understanding of low-dimensional chaotic systems,
 less is known about spatially extended systems
 whose number of active DOFs is large or infinite.
This is partly because of difficulty in treating concepts such as measures
 for infinite-dimensional dynamical systems in a mathematically proper way
 \cite{MathReviews-1},
 and partly because of computational complexity;
 e.g. with regarding to the UPO expansion,
 not only does the number of UPOs grow exponentially with increasing DOFs,
 but even finding one UPO becomes much more laboring.
However, the number of DOFs one can numerically
 investigate increases gradually,
 which makes it possible for various theoretical concepts and methods
 to be extended and applied to high-dimensional chaos
 \cite{NumericWorks-1,Politi_Torcini-1}.
It leads to reveal several suggestive properties
 intrinsic to spatially extended systems, which have been reported recently.
For example, it was found that
 one can reproduce macroscopic quantities of turbulence only from a single UPO
 \cite{1UPO-1,Kawasaki_Sasa-1}.

One of the most striking manifestations of high dimensionality is
 the occurrence of phase transitions.
In the case of coupled map lattices (CMLs),
 i.e. lattices of interacting dynamical systems whose time evolution is
 defined by a map, 
 logistic CMLs are known to display nontrivial collective behavior
 which cannot be observed in equilibrium systems,
 and transitions between two types of collective behavior
 can be regarded as phase transitions
 \cite{Chate_Manneville-1,Marcq_Chate_Manneville-2}.
Another interesting example of non-equilibrium phase transitions
 is 2-dimensional CMLs which exhibit
 a continuous phase transition similar to that of the Ising model
 \cite{Miller_Huse-1,Marcq_Chate_Manneville-1}.
The existence of a new universality class was numerically shown for such CMLs
 with synchronously updating rules,
 while those asynchronously updated belong to the 2D Ising universality class
 \cite{Marcq_Chate_Manneville-1}.
Recent studies suggest that the Ising-like transitions of synchronous CMLs
 and the onset of the non-trivial collective behavior of a logistic CML
 belong to the same universality class,
 i.e. the non-Ising class \cite{Marcq_Chate_Manneville-2}.

Although many interesting properties of non-equilibrium phase transitions
 have been found out,
 there seems nevertheless no consensus
 on the usage of the term ``phase transition'' in dynamical systems.
Theoretically, it can be defined as a qualitative change
 in the statistical behavior of typical orbits in a single mixing attractor
 which does not change topologically
 \cite{Gielis_MacKay-1,Just_Schmuser-1,Blank_Bunimovich-1,Bardet_Keller-1},
 by which we exclude bifurcations coming up
 even in finite-dimensional dynamical systems.
For the definition of ``qualitative change,''
 the analogy with that in equilibrium phase transitions is used.
There are two complementary manners of characterizing equilibrium transitions
 \cite{vanEnter_etal-1} : 
 one is after Ehrenfest,
% which is based on the singularity of the free energy
 where $n$-th order phase transitions are identified
 as divergence or discontinuity of some $n$-th derivative of the free energy.
The other is after Gibbs,
% which is based on the number of
 where first-order phase transitions correspond to a change in the number of
 the pure Gibbs measures, or macrostates.
Analogues of the latter have been adopted in the context of dynamical systems
 since no free energy appears useful:
 if we consider interaction in a formal Hamiltonian
 on the space-time configuration space and define a free energy from it,
 then the analyticity of the free energy is a very delicate problem
 \cite{Bricmont_Kupiainen-1,MathReviews-1}
 and too complicated to relate to phase transitions.
On the other hand, if we define a free energy
 from purely probabilistic measure approach as we will see below,
 then it is identically zero and thus analytic
 in the whole parameter region, even at criticality,
 due to a strong constraint which comes from a normalization of the measure
 \cite{Lebowitz_etal-1}.
Therefore, transitions have been defined in the Gibbsian sense
 \cite{Gielis_MacKay-1,Blank_Bunimovich-1,Bardet_Keller-1},
 that is via a change of the number of natural measures,
 which corresponds to first order transitions.
This definition, however, cannot characterize higher order transitions
 as definitively, so we have to make use of more subtle phenomena,
 such as spontaneous symmetry breaking, divergence of a correlation length,
 formation of an infinite cluster, and so on.
It is true that they are closely related to phase transitions,
 but would not prescribe them as quantitatively as equilibrium counterparts do.
Thus, it is desirable to develop another way
 to characterize phase transitions in extended chaos,
 including higher order ones.

Another issue involved with phase transitions in extended chaotic systems
 is absence of macroscopic ``temperature'' which controls the systems.
Some locally defined parameters
 such as coupling strength and diffusion coefficient
 have been used as \textit{ad hoc} substitutes for temperature
 (e.g. \cite{Marcq_Chate_Manneville-2,Marcq_Chate_Manneville-1}),
 while its theoretical grounds remain to be clarified.
This treatment is based on an assumption
 that such local parameters are direct barometers of macroscopic properties.
This is however not at all trivial,
 as we can see for example from studies of effective temperature
 in non-equilibrium systems \cite{EffectiveTemperature-1}.
Although we can argue the issue to some extent
 by renormalization group approach,
 it cannot deal with concrete systems.
Therefore, it is desirable to have a basis
 on which we can connect locally defined time evolution rule of a system
 to macroscopic properties.

In the present paper, we characterize phase transitions
 in extended chaotic systems, namely CMLs,
 including both equilibrium and non-equilibrium ones.
The periodic orbit expansion and the thermodynamic formalism
 are adapted for such systems, by which
 the relation between q-phase transitions and (actual) phase transitions
 is investigated.
The main outcomes are that
 (1) a rather quantitative way to define non-equilibrium phase transitions
 with their order in the Ehrenfest's sense is proposed,
 and (2) a basis is obtained
 on which microscopic control parameters can be handled similarly
 to temperature in some cases involved with phase transitions.
Note that it is not the aim of this paper
 to give a mathematically rigorous argument,
 which is often highly delicate in this field
 \cite{MathReviews-1,Bricmont_Kupiainen-1}
 and may limit an attainable conclusion.
Instead, we shall devote ourselves to obtain a physically plausible picture.

This paper is organized as follows :
 we first review the idea of UPO ensemble
 \cite{Kawasaki_Sasa-1,UPO_Expansion-1}
 (Sec. \ref{sec:2.a}), and on its basis
 the thermodynamic formalism is formulated
 to deal with dynamical averages and fluctuations
 of some macroscopic quantity in chaotic systems with many DOFs
 (Sec. \ref{sec:2.b}).
A corresponding partition function and topological pressure,
 or ``free energy'', are defined
 and the moments are obtained as its differential coefficients.
Then we apply it to a 1D Bernoulli CML (Sec. \ref{sec:3})
 which can be regarded as a deterministic model
 of the 1D Ising model as is summarized in Sec. \ref{sec:3.a}.
After we mention the computational procedure for the thermodynamic formalism
 (Sec. \ref{sec:3.b}),
we show that an anomalous distribution of UPOs
 exists in such a system with phase transitions
 (marginal transitions in this example),
 which can be regarded as a seed of the Ising transition (Sec. \ref{sec:3.c}).
The seed is embodied as a q-phase transition.
Another example is a 1D repelling CML
 which exhibits a non-marginal transition (Sec. \ref{sec:4}).
This is a solvable case, hence we can explicitly see
 the relation between phase transitions and q-phase transitions.
Section \ref{sec:5} is assigned to the discussion and conclusion.
Note that the terminologies ``phase transition''
 and ``q-phase transition'' are specifically discriminated
 throughout this paper.

\section{Thermodynamic formalism\protect\\ for extended systems}  \label{sec:2}

\subsection{UPO ensemble}\label{sec:2.a}

First, we review the concept of UPO ensemble
 \cite{Kawasaki_Sasa-1,UPO_Expansion-1},
 on which the following thermodynamic formalism is based.
This and the next subsections are assigned to show
 the grounds for our arguments in the rest of sections
 and the range of their applications.

Consider a dynamical system with discrete time,
 $\px^{t+1} = \pF(\px^t), ~~\px^{t} \equiv [ x_0^t , x_1^t , \cdots , x_{N-1}^t ]$,
 where $N$ denotes the number of DOFs and is large.
Our goal for the time being is to obtain the dynamical average
 of an arbitrary macroscopic observable $A(\px)$,
 which is defined as a function of the dynamical variable $\px$.
Here the term ``macroscopic observable'' represents a quantity
 obtained by taking the average over the DOFs of the system.
Suppose the system is ergodic, the long-time average
 $\expct{A}_\text{time} \equiv \lim_{n\to\infty}(1/n)\sum_{t=0}^{n-1} A(\px^t)$
 is equal to the phase space average
 $\expct{A}_\mu \equiv \int A(\px) \mu(\rd \px)$
 for almost all initial conditions $\px^0$,
 where $\mu$ denotes the SRB measure, or the natural invariant measure.
For mixing and hyperbolic systems, the following relation between
 the natural invariant measure of a subset $S$ and UPOs holds
 \cite{UPO_Expansion-1}
\begin{equation}
 \mu(S) = \lim_{p \to \infty} \sum_{\{[\px];\px^0 \in S\}} \e^{-pN\lambda([\px])}.  \label{eq:2.1}
\end{equation}
Here, $[\px] \equiv \px^0 \px^1 \cdots \px^{p-1}$
 indicates an UPO of period $p$
 and therefore the sum in \eqref{eq:2.1} is taken over all the UPOs
 of period $p$ which start from $S$ and return to it.
$\lambda([\px])$ is a positive Lyapunov exponent per 1 DOF
\begin{equation}
 \lambda([\px]) \equiv \frac{1}{N}\sum_{i=0}^{N-1} \lambda^+_i([\px]),  \label{eq:2.2}
\end{equation}
 where $\{ \lambda^+_i([\px]) \}$ denotes a set of positive exponents
 of the UPO $[\px]$.
Note that we sometimes call $\lambda([\px])$ simply ``Lyapunov exponent''
 as long as it does not cause any confusion.
From \eqref{eq:2.1}, $\e^{-pN\lambda([\px])}$ can be regarded as
 the probability measure of the UPO ensemble.

For fixed $p$ longer than the time required for mixing,
 denoted by $\tau_\text{mixing}$, \eqref{eq:2.1} holds approximately,
 so that $\expct{A}_\mu$ can be estimated from the ensemble average
 $\expct{A(\px^0)}_\text{UPO} \equiv \sum_{[\px]} A(\px^0) \e^{-pN\lambda([\px])} \simeq \int A(\px^0) \mu(\rd\px^0) = \expct{A}_\mu$.
Moreover, since $\lambda([\px])$ is invariant under cyclic permutation
 of $[\px] = \px^0 \px^1 \cdots \px^{p-1}$,
 $\expct{A(\px^t)}_\text{UPO} \simeq \expct{A}_\mu$ also stands,
 and consequently for the average along an UPO
\begin{equation}
 A([\px]) \equiv \frac{1}{p} \sum_{t=0}^{p-1} A(\px^t),  \label{eq:2.3}
\end{equation}
 the following relation holds if $p \gtrsim \tau_\text{mixing}$
\begin{equation}
 \expct{A([\px])}_\text{UPO} \equiv \sum_{[\px]} A([\px]) \e^{-pN\lambda([\px])} \simeq \expct{A}_\mu .  \label{eq:2.4}
\end{equation}
Note that the period $p$ required to make \eqref{eq:2.4} converge
 can be shorter than $\tau_\text{mixing}$ if $N$ is sufficiently large,
 thanks to the law of large numbers \cite{Kawasaki-PC}.
The estimation of $\expct{A}_\mu$ from $A([\px])$ defined by \eqref{eq:2.3}
 is preferable to that from $A(\px^t)$, because its variance
\begin{align}
 \sigma(A([\px]))_\text{UPO}^2
 &\equiv \Expct{ \big( A([\px]) - \expct{A([\px])}_\text{UPO} \big)^2 }_\text{UPO}  \notag \\
 &= \frac{1}{p} \sum_{\tau=0}^{p-1} \Cumulant{A(\px^t)}{A(\px^{t+\tau})}_\text{UPO}  \label{eq:2.5}
\end{align}
 does not exceed the variance of $A(\px^t)$, 
 $\sigma(A(\px^t))_\text{UPO}^2 \equiv \Cumulant{A(\px^t)}{A(\px^t)}_\text{UPO} \simeq \sigma(A)_\mu^2$.
If we assume that the autocorrelation function decays exponentially
 $\Cumulant{A(\px^t)}{A(\px^{t+\tau})}_\text{UPO} \sim \e^{-\tau/\tau_0}$
 with the correlation time
 $1 \ll \tau_0 \ll p/2$, the ratio of the 2 variances is
 $\sigma(A([\px]))_\text{UPO}^2 \simeq (2\tau_0/p) \sigma(A)_\mu^2$.

As is shown above, the phase space average of the macroscopic quantity $A(\px)$
 and the lower bound of its fluctuation are obtained
 from the UPO ensemble treatment, which are approximately equal to those
 time averaged along typical orbits.

\subsection{Thermodynamic formalism}  \label{sec:2.b}

In this subsection, we introduce an appropriate partition function
 to deal with dynamical averages and fluctuations in extended systems,
 that is
\begin{equation}
 Z_{q,\beta} \equiv \sum_{[\px]} \e^{-pN\left[ q\lambda([\px]) + \beta A([\px]) \right]},  \label{eq:2.6}
\end{equation}
 where the sum is taken over all of the UPOs whose period is $p$.
The summation without the second term in the exponential
 represents the Lyapunov partition function \cite{Lyap.Part.Func.-1}.
Variables $q$ and $\beta$ inserted in \eqref{eq:2.6} are auxiliary ones,
 which can be regarded as inverse temperature mathematically,
 but of no particular physical significance.
However, since we can change the dominant terms in the sum of \eqref{eq:2.6}
 by varying $q$ and $\beta$,
 they play essential roles in the following argument.
The real system corresponds to $(q,\beta)=(1,0)$,
 where the summands in the partition function coincide with
 the probability measures of the UPO ensemble,
 hence we call it \textit{physical situation} hereafter.
Note that the partition function \pref{eq:2.6} is similar to
 that introduced by Fujisaka and Inoue \cite{Fujisaka_Inoue-1},
 but here we explicitly consider the scaling dependence
 on the number of DOFs $N$ as well as the period $p$
 in order to argue phase transitions.

The relation to the space-time Gibbs measure should be also referred to.
The space-time measure is often introduced as a measure
 of refinement elements (so-called cylinder) under the symbolization
 \cite{Gielis_MacKay-1,Just_Schmuser-1}
 and the accompanying partition function is
 equal to that in \eqref{eq:2.6} with $(q,\beta)=(1,0)$.
The partition function mentioned in this paper is constituted
 by adding an observable $A([\px])$ to the argument of the exponential term
 and introducing ``temperature'' parameters
 based on the thermodynamic formalism.
What is essential in the concept of the space-time measure is that
 we can consider its configuration space to be
 the $(d+1)$-dimensional space-time
 comprising the $d$-dimensional space and the 1-dimensional time,
 which remains valid after the extension.
That is to say, the partition function defined here 
 are natural extensions of those of the space-time measure,
 which means we can exploit plentiful knowledge
 in the equilibrium statistical mechanics
 to the problem of spatio-temporal chaos.

The corresponding ``free energy'',
 which is called the topological pressure in the context of dynamical systems
 \cite{Ruelle-1,Beck-1}
 but we call it here the generalized Massieu function (GMF),
 is defined by
\begin{equation}
 \Psi(q,\beta) \equiv -\frac{1}{pN} \ln Z_{q,\beta}.  \label{eq:2.7}
\end{equation}
Note that the sign of \eqref{eq:2.7} is set
 opposite to the conventional definition of the topological pressure
 so as to mention a minimum principle of it.
Since both $\lambda([\px])$ and $A([\px])$ can be regarded
 as intensive densities per unit time and one DOF,
 they remain finite in the limit $p,N \to \infty$
 and the GMF $\Psi(q,\beta)$ is expected to converge in that limit.
Especially, since the measure of the whole phase space is one,
 \eqsref{eq:2.1}, \pref{eq:2.6}, and \pref{eq:2.7} yield
\begin{equation}
 \lim_{p \to \infty} \Psi(1,0)=0.  \label{eq:2.8}
\end{equation}
This constraint must be satisfied at the physical situation
 regardless of values of control parameters.
This fact prevents us from defining phase transitions
 just by the sigularity of the ``free energy'' with respect to parameters.
We shall see, however, that
 by introducing a generalized probability measure in \eqref{eq:2.6},
 we make room for \textit{the singularity with respect to $q$ and $\beta$},
 which is called q-phase transition \cite{Hata_etal-1},
 and thus we are in fact able to relate actual phase transitions
 (with respect to parameters)
 to the singularity of the ``free energy.''
This point will be clarified in Sec. \ref{sec:3.c}.

The ensemble average and fluctuation of $A([\px])$
 defined by \eqsref{eq:2.4} and \pref{eq:2.5} respectively
 are obtained from the differential coefficients of the GMF by
\begin{align}
 &\expct{A}_\mu \simeq \expct{A([\px])}_\text{UPO} = \prt{\Psi}{\beta}(1,0),  \label{eq:2.9} \\
 &\sigma(A)_\mu^2 \gtrsim \sigma(A([\px]))_\text{UPO}^2 = -\frac{1}{pN} \prts{\Psi}{\beta}{2}(1,0),  \label{eq:2.10}
\end{align}
 where the UPO average $\expct{A([\px])}_\text{UPO}$ is redefined as
 $\expct{A([\px])}_\text{UPO} \equiv \sum_{[\px]} A([\px]) \e^{-pN\lambda([\px])} / \sum_{[\px]} \e^{-pN\lambda([\px])}$ in order to moderate
 errors due to the finite-size effect.
These relations are completely analogous
% in a mathematical sense
 to counterparts of the canonical statistical mechanics
 and therefore all moments of $A([\px])$ can be obtained
 by differentiating the GMF up to the requisite order.
The average and variance of
 the positive finite-time Lyapunov exponent per 1 DOF
 can be also acquired
 without replacing the definition of $A(\px)$ by them;
\begin{align}
 &\expct{\lambda([\px])}_\mu \simeq \expct{\lambda([\px])}_\text{UPO} = \prt{\Psi}{q}(1,0),  \label{eq:2.11} \\
 &\sigma(\lambda([\px]))_\mu^2 \simeq \sigma(\lambda([\px]))_\text{UPO}^2 = -\frac{1}{pN} \prts{\Psi}{q}{2}(1,0).  \label{eq:2.12}
\end{align}
The positive (infinite-time) Lyapunov exponent per 1 DOF can be obtained
 by taking a limit $p \to \infty$ in \eqref{eq:2.11}.
Moreover, the equalities and inequality \pref{eq:2.9}-\pref{eq:2.12}
 hold precisely in that limit.
 They are expected to be good approximations for a finite period $p$,
 at least if $p \gtrsim \tau_\text{mixing}$,
 as is mentioned in the previous subsection.
%The generalized entropy $K_q$ per 1 DOF is also obtained from the GMF by
%\begin{equation}
% K_q = \frac{1}{q-1} \lim_{p\to\infty} \Psi(q,0),  \label{eq:2.13}
%\end{equation}
% which is investigated by Politi and Torcini
% for a lattice of coupled H\'{e}non maps \cite{Politi_Torcini-1}.

\begin{figure}[t]
 \includegraphics[clip]{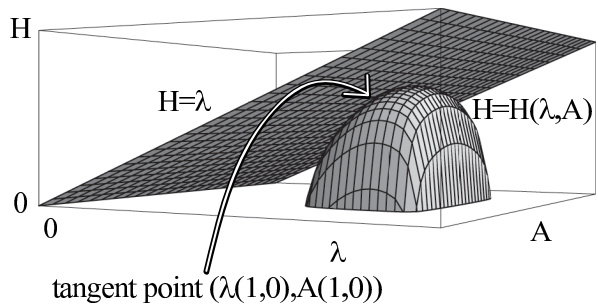}
 \caption{A schematic view of an expected form of the topological entropy spectrum $H(\lambda,A)$.}
 \label{fig:1}
\end{figure}

Now we consider the statistics of UPOs,
 namely the distribution of UPOs
 with respect to their macroscopic properties.
Let $\Omega (\lambda,A) \rd \lambda \rd A$ denote the number of UPOs
 whose positive Lyapunov exponent $\lambda([\px])$
 and macroscopic quantity $A([\px])$ are within the range of
 $\lambda$-$\lambda+\rd\lambda$ and $A$-$A+\rd A$ respectively.
Suppose the system is homogeneous, in other words
 the system consists of identical DOFs
 and thus it can be viewed as an ensemble of smaller coupled subsystems,
 we can assume the following functional form of 
 $\Omega (\lambda,A) \rd \lambda \rd A$:
\begin{equation}
 \Omega(\lambda,A)\rd\lambda \rd A \sim \rho(\lambda,A;p,N) \e^{pNH(\lambda,A)} \rd\lambda \rd A.  \label{eq:2.14}
\end{equation}
Here $\rho(\lambda,A;p,N)$ is a ``coefficient'' into which
 all factors are pushed whose dependence on $pN$ is not exponential.
$H(\lambda,A)$ is a concave function,
 which is considered to be a topological entropy per 1 DOF
 under the restriction of
 $\lambda([\px]) \in [\lambda,\lambda+\rd\lambda]$
 and $A([\px]) \in [A,A+\rd A]$.
Roughly speaking, the expression \pref{eq:2.14} is justified
 by the large deviation theorem because both $\lambda([\px])$ and $A([\px])$
 can be regarded as the averages over $pN$ variables
 which correlate to each other with a specific correlation time and length.
By making use of the distribution function \pref{eq:2.14}
 to calculate the partition function \pref{eq:2.6},
 we obtain
\begin{align}
 Z_{q,\beta}
 &= \int \e^{-pN(q\lambda + \beta A )} \Omega(\lambda,A) \rd \lambda \rd A  \notag \\
 &\sim \int \rho(\lambda,A;p,N) \e^{-pN[q\lambda + \beta A -H(\lambda,A)]} \rd \lambda \rd A.  \label{eq:2.15}
\end{align}
If the product of the period of the UPOs and the number of the DOFs, $pN$,
 is sufficiently large,
 the saddle point approximation is applicable, that is,
 only the vicinity of the point $(\lambda,A)=(\lambda(q,\beta),A(q,\beta))$
 where the integrand has a maximum contributes to the integral \pref{eq:2.15}.
The conditions imposed on $\lambda(q,\beta),A(q,\beta)$ are
\begin{subequations}
\begin{align}
 &\prt{H}{\lambda}=q,~~~~\prt{H}{A}=\beta,  \label{eq:2.16a} \\
 &\prts{H}{\lambda}{2}+\prts{H}{A}{2}<0,~~~~
 \det \begin{pmatrix} \prts{H}{\lambda}{2} & \frac{\p^2 H}{\p\lambda \p A} \\ \frac{\p^2 H}{\p\lambda\p A} & \prts{H}{A}{2} \end{pmatrix}>0,  \label{eq:2.16b}
\end{align}  \label{eq:2.16}
\end{subequations}
 where all differential coefficients of $H(\lambda,A)$
 are evaluated at $(\lambda,A)=(\lambda(q,\beta),A(q,\beta))$.
Using the saddle point approximation to \eqref{eq:2.15}
 and substituting it for \eqref{eq:2.7}, we obtain
\begin{equation}
 \Psi(q,\beta)
 \simeq \min_{\lambda,A} \left[ q\lambda + \beta A -H(\lambda,A) \right],  \label{eq:2.17}
\end{equation}
 or
\begin{equation}
 \Psi(q,\beta) \simeq q\lambda(q,\beta) + \beta A(q,\beta) - H(\lambda(q,\beta),A(q,\beta)).  \label{eq:2.18}
\end{equation}
These equations hold rigorously in the limit $p,N \to \infty$.
Equation \pref{eq:2.17} can be regarded as a principle of minimum free energy
 in the sense that $\lambda$ and $A$ which are dominant
 in the partition function \pref{eq:2.15} are selected out
 to minimize the corresponding GMF.
The relation \pref{eq:2.18} accompanied by \eqref{eq:2.16a}
 is the Legendre transformation and thus
\begin{equation}
 \lambda(q,\beta) \simeq \prt{\Psi}{q}(q,\beta),~~~~
 A(q,\beta) \simeq \prt{\Psi}{\beta}(q,\beta),  \label{eq:2.19}
\end{equation}
 which are obtained by differentiating \eqref{eq:2.18} by $q$ or $\beta$.
The comparison of \eqref{eq:2.19} with \eqref{eq:2.9} and \pref{eq:2.11}
 yields the relations $\expct{A([\px])}_\text{UPO} \simeq A(1,0),\expct{\lambda([\px])}_\text{UPO} \simeq \lambda(1,0)$, which appear to be natural
 because the right hand sides represent the dominant $A$ and $\lambda$
 at the physical situation.
In addition, the concavity of $H(\lambda,A)$ and
 the relations \pref{eq:2.16a}, \pref{eq:2.18}, and \pref{eq:2.8}
 yield a vision of a general form
 of the function $H(\lambda,A)$.
It is expected to be tangent to a plane $H=\lambda$ at the physical point
 $(q,\beta)=(1,0)$, as illustrated schematically in Fig. \ref{fig:1},
 and the tangent point represents a state which is observed physically.

\section{Analysis of the 1D Bernoulli CML}  \label{sec:3}

\subsection{Model}  \label{sec:3.a}

The map we first analyze is a Bernoulli CML,
 whose 2D version was originally proposed
 by Sakaguchi \cite{Sakaguchi-1}
 and its 1D version was introduced later
 by Kawasaki and Sasa \cite{Kawasaki_Sasa-1}.
In the present work, we investigate the 1D model, which we describe below.
\begin{figure}[t]
%WinTpicVersion3.08
\unitlength 0.1in
\begin{picture}( 26.9000, 27.9000)(  1.2000,-28.5000)
% BOX 2 0 3 0
% 2 500 500 2500 2500
% 
\special{pn 8}%
\special{pa 500 500}%
\special{pa 2500 500}%
\special{pa 2500 2500}%
\special{pa 500 2500}%
\special{pa 500 500}%
\special{fp}%
% VECTOR 2 0 3 0
% 4 250 1500 2750 1500 1500 2750 1500 250
% 
\special{pn 8}%
\special{pa 250 1500}%
\special{pa 2750 1500}%
\special{fp}%
\special{sh 1}%
\special{pa 2750 1500}%
\special{pa 2684 1480}%
\special{pa 2698 1500}%
\special{pa 2684 1520}%
\special{pa 2750 1500}%
\special{fp}%
\special{pa 1500 2750}%
\special{pa 1500 250}%
\special{fp}%
\special{sh 1}%
\special{pa 1500 250}%
\special{pa 1480 318}%
\special{pa 1500 304}%
\special{pa 1520 318}%
\special{pa 1500 250}%
\special{fp}%
% STR 2 0 3 0
% 3 2810 1520 2810 1620 2 0
% $x_i^t$
\put(28.1000,-16.2000){\makebox(0,0)[lb]{$x_i^t$}}%
% STR 2 0 3 0
% 3 1290 130 1290 230 2 0
% $x_i^{t+1}$
\put(12.9000,-2.3000){\makebox(0,0)[lb]{$x_i^{t+1}$}}%
% LINE 2 0 3 0
% 4 500 2500 1900 500 1900 2500 2500 500
% 
\special{pn 8}%
\special{pa 500 2500}%
\special{pa 1900 500}%
\special{fp}%
\special{pa 1900 2500}%
\special{pa 2500 500}%
\special{fp}%
% LINE 2 1 3 0
% 2 1900 500 1900 2500
% 
\special{pn 8}%
\special{pa 1900 500}%
\special{pa 1900 2500}%
\special{da 0.070}%
% STR 2 0 3 0
% 3 1760 1370 1760 1470 2 0
% $\Delta_i^t$
\put(17.0000,-14.7000){\makebox(0,0)[lb]{$\Delta_i^t$}}%
% STR 2 0 3 0
% 3 480 1380 480 1480 3 0
% $-1$
\put(4.8000,-14.8000){\makebox(0,0)[rb]{$-1$}}%
% STR 2 0 3 0
% 3 2540 1370 2540 1470 2 0
% $+1$
\put(25.4000,-14.7000){\makebox(0,0)[lb]{$+1$}}%
% STR 2 0 3 0
% 3 1530 360 1530 460 2 0
% $+1$
\put(15.3000,-4.6000){\makebox(0,0)[lb]{$+1$}}%
% STR 2 0 3 0
% 3 1540 2470 1540 2570 1 0
% $-1$
\put(15.4000,-25.7000){\makebox(0,0)[lt]{$-1$}}%
% STR 2 0 3 0
% 3 1060 2610 1060 2710 5 0
% $s_i^t=+1$
\put(10.6000,-27.1000){\makebox(0,0){$s_i^t=+1$}}%
% STR 2 0 3 0
% 3 2220 2610 2220 2710 5 0
% $s_i^t=-1$
\put(22.2000,-27.1000){\makebox(0,0){$s_i^t=-1$}}%
% VECTOR 2 0 3 0
% 2 1200 2800 500 2800
% 
\special{pn 8}%
\special{pa 1200 2800}%
\special{pa 500 2800}%
\special{fp}%
\special{sh 1}%
\special{pa 500 2800}%
\special{pa 568 2820}%
\special{pa 554 2800}%
\special{pa 568 2780}%
\special{pa 500 2800}%
\special{fp}%
% VECTOR 2 0 3 0
% 2 1200 2800 1890 2800
% 
\special{pn 8}%
\special{pa 1200 2800}%
\special{pa 1890 2800}%
\special{fp}%
\special{sh 1}%
\special{pa 1890 2800}%
\special{pa 1824 2780}%
\special{pa 1838 2800}%
\special{pa 1824 2820}%
\special{pa 1890 2800}%
\special{fp}%
% VECTOR 2 0 3 0
% 2 2200 2800 1910 2800
% 
\special{pn 8}%
\special{pa 2200 2800}%
\special{pa 1910 2800}%
\special{fp}%
\special{sh 1}%
\special{pa 1910 2800}%
\special{pa 1978 2820}%
\special{pa 1964 2800}%
\special{pa 1978 2780}%
\special{pa 1910 2800}%
\special{fp}%
% VECTOR 2 0 3 0
% 2 2200 2800 2500 2800
% 
\special{pn 8}%
\special{pa 2200 2800}%
\special{pa 2500 2800}%
\special{fp}%
\special{sh 1}%
\special{pa 2500 2800}%
\special{pa 2434 2780}%
\special{pa 2448 2800}%
\special{pa 2434 2820}%
\special{pa 2500 2800}%
\special{fp}%
% LINE 2 5 3 0
% 2 500 2850 2500 2850
% 
\special{pn 8}%
\special{pa 500 2850}%
\special{pa 2500 2850}%
\special{ip}%
\end{picture}%
 \caption{The local Bernoulli map at a site $i$. The time evolution of $\Delta_i^t$ is defined in \eqref{eq:3.2}.}
 \label{fig:2}
\end{figure}
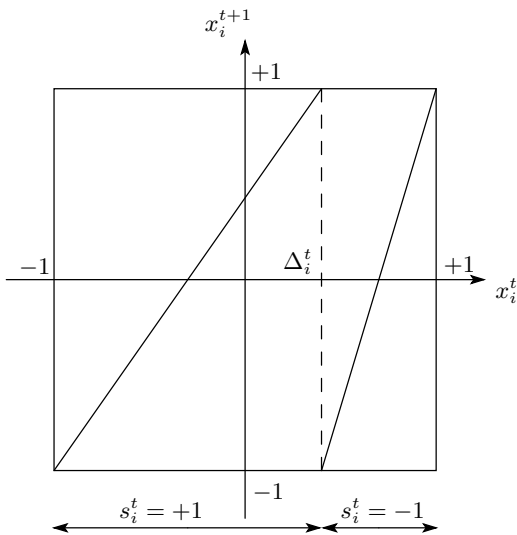%

Consider a 1-dimensional lattice
 which consists of $N$ lattice points $i=0,1,\cdots,N-1$.
Dynamical variables $(x_i,\Delta_i) \in [-1,1] \times [-1,1]$ are assigned
 to each site $i$, and in addition, a ``spin'' variable $s_i$ is defined as
\begin{equation}
 s_i \equiv \begin{cases} +1 & \text{if $-1 \leq x_i < \Delta_i$}, \\ -1 & \text{if $\Delta_i \leq x_i \leq 1$}. \end{cases}  \label{eq:3.1}
\end{equation}
With this spin, the time evolution of $(x_i^t,\Delta_i^t)$ is written as
\begin{equation}
 \begin{cases}
  \begin{cases}
   x_i^{t+1} = f(x_i^t;\Delta_i^t) \equiv \dfrac{2(x_i^t + s_i^t)}{1+s_i^t \Delta_i^t} - s_i^t, \\
   \Delta_i^{t+1} = \tanh \left[ \dfrac{k}{2}(s_{i-1}^t + s_{i+1}^t) \right],
  \end{cases} & \hspace{-15pt}\text{for odd $(i-t)$}, \\
  \begin{cases} 
   x_i^{t+1} = x_i^t, \\
   \Delta_i^{t+1} = \Delta_i^t,
  \end{cases} & \hspace{-15pt}\text{for even $(i-t)$},
 \end{cases}  \label{eq:3.2}
\end{equation}
 with periodic boundary condition $s_N^t=s_0^t$.
The updating is done alternately with respect to the parity
 of the site number $i$, that is to say, sites with odd $i$ are
 updated at even time $t$ while those with even $i$
 are renewed at odd $t$.
The total number of the sites $N$ is supposed to be even
 in order that the alternately updating rule is compatible
 with the periodic boundary condition.
$f(x_i^t;\Delta_i^t)$ is a Bernoulli map,
 illustrated in Fig. \ref{fig:2}.
As can be seen from \eqref{eq:3.2} and Fig. \ref{fig:2},
 $\Delta_i^t$ is a discrete variable and behaves as a dynamical parameter
 which describes the interaction between nearest-neighbor sites.
Therefore, we consider only $x_i^t$ to be a dynamical variable
 and apply the formalism stated in the previous section.
The magnitude and the tendency of the interaction are determined
 by the absolute value of $k$ and its sign, respectively.
For positive (negative) $k$,
 $\Delta_i^t$ moves in the direction so as to
 make it more probable that the spin $s_i^t$ becomes
 parallel (antiparallel) to the neighboring spins,
 hence the interaction is ferromagnetic (antiferromagnetic).

The Bernoulli CML has several remarkable properties,
 as demonstrated by preceding studies \cite{Kawasaki_Sasa-1,Sakaguchi-1},
 which should be pointed out here.
First, the dynamics can be expressed in terms of the symbolic dynamics
 with symbols $\ps \equiv \{ s_i \}_{i=0}^{N-1}$.
In other words, the partition $\{ U_\ps \}$ of the phase space,
 whose element $U_\ps$ corresponds to a spin configuration $\ps$,
 is generating and thus every orbit is specified by an infinite sequence
 of symbols $\ps^0 \ps^1 \ps^2 \cdots$.
Especially, note that every UPO has a one-to-one correspondence
 to a finite length permutation $[\ps] \equiv \ps^0 \ps^1 \cdots \ps^{p-1}$.
The most significant feature of the Bernoulli CML is that
 it respects a detailed balance and
 the resulting probability measure of a subset $U_\ps$ coincides
 with the canonical distribution of the 1D Ising model
 \cite{Kawasaki_Sasa-1,Sakaguchi-1}, namely
\begin{equation}
 \mu(U_\ps) \propto \exp \left( \frac{k}{2} \sum_{i=0}^{N-1} s_i s_{i+1} \right) .   \label{eq:3.3}
\end{equation}
Therefore the Bernoulli CML can be regarded as a deterministic model
 of the 1D Ising (anti-)ferromagnetism in its equilibrium state
 and the interaction parameter $k$ corresponds to the inverse temperature.
Since the marginal phase transition occurs in the 1D Ising model
 at the zero temperature limit,
 this Bernoulli CML shows a transition
 in the strong interaction limit $|k| \to \infty$.

\subsection{Application of the thermodynamic formalism}  \label{sec:3.b}

\begin{figure}[b]
 \includegraphics[clip]{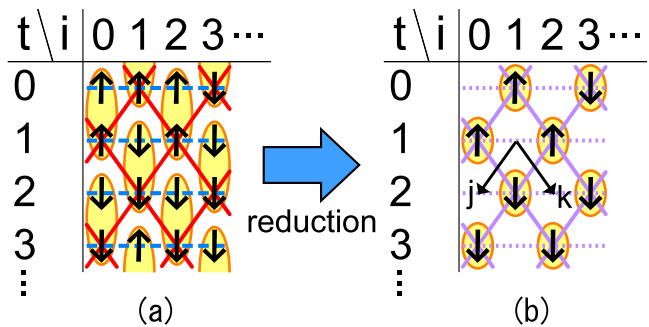}
 \caption{(Color online) A schematic illustration of the spin reduction from an UPO  $[\ps]$ to the corresponding 2-dimensional array $\{ s_{j,k} \}$. (a) The spin configuration of an UPO $[\ps]$. Spins form clusters of length $2$ in the time direction under the updating rule of \eqref{eq:3.2}. The red solid line and the blue broken line indicate spin interaction which comes from the Lyapunov exponent $\lambda([\ps])$ and the observable $A([\ps])$, respectively, in \eqref{eq:2.6}. (b) The 2-dimensional spin array $\{ s_{j,k} \}$ obtained by the reduction, by which each cluster is reduced to a single spin located at the odd $(i-t)$. The 2 kinds of spin interaction turn into helical ferromagnetic part (purple solid line) and spatial antiferromagnetic part (purple dotted line) suppose both $(\beta+kq/2)$ and $q$ are positive.}
 \label{fig:3}
\end{figure}

The thermodynamic formalism in Sec. \ref{sec:2.b} is made use of
 to analyze it.
We adopt the Ising interaction energy per 1 spin for a macroscopic quantity
\begin{equation}
 A(\ps) \equiv -\frac{1}{N} \sum_{i=0}^{N-1} s_i s_{i+1}.  \label{eq:3.4}
\end{equation}
Substituting it and the Lyapunov exponent given from the slope of the function
 $f(x_i^t;\Delta_i^t)$ into \eqref{eq:2.6},
 we can obtain the following expression of the partition function
\begin{widetext}
\begin{equation}
 Z_{q,\beta}
 = \sum_{\{ s_{j,k} \}} \exp \[ \( \beta+\frac{kq}{2} \) \sum_\text{n.n.} s_{j,k} s_{j',k'} -\frac{q}{2} \( \ln \cosh k \) \sum_{j,k} s_{j-1,k} s_{j,k-1} -\frac{pNq}{4} \ln (4\cosh k) \],  \label{eq:3.5}
% = \( \frac{1}{4\cosh k} \)^{pNq/4} \sum_{\{ s_{j,k} \}} \exp \[ \( \beta+\frac{kq}{2} \) \sum_\text{n.n.} s_{j,k} s_{j',k'} -\frac{q}{2} \( \ln \cosh k \) \sum_{j,k} s_{j-1,k} s_{j,k-1} \],  \label{eq:3.5}
\end{equation}
\end{widetext}
 where a space-time configuration $[\ps]$ of $N \times p$ symbols is reduced
 to a 2-dimensional array $\{ s_{j,k} \}$ of $pN/2$ spins
 by exploiting the constraint $s_i^{t+1} = s_i^t$ for even $(i-t)$,
 which is outlined in Fig. \ref{fig:3},
 and $\sum_\text{n.n.}$ indicates a summation
 over all pairs of neighboring spins after the spin reduction.
Equation \pref{eq:3.5} shows that,
 for positive $(\beta+kq/2)$ and $q$,
 the interaction between spins comprises
 helical ferromagnetic part and spatial antiferromagnetic part.

\begin{figure*}
 \includegraphics[clip]{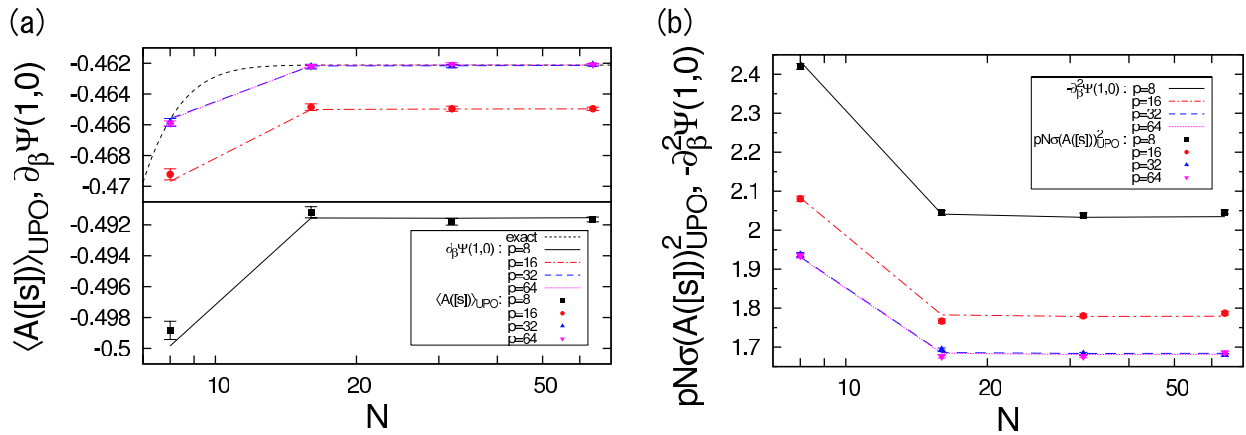}
 \caption{(Color online) Demonstration of \eqsref{eq:2.9} and \pref{eq:2.10} for the 1D Bernoulli CML with $k=1$, by means of Monte Carlo calculations. Lines correspond to (a) mean $\prt{\Psi}{\beta}(1,0)$ and (b) standard deviation $-\prts{\Psi}{\beta}{2}(1,0)$ at $N=8,16,32,64$ evaluated via the GMF $\Psi(q,\beta)$, which is obtained by averaging results of 400 independent Monte Carlo runs with $100,000$ samples after $100$ steps of transients. The range of errors, estimated from standard deviation among the independent runs, is less than $10^{-4}$ for (a) and $10^{-2}$ for (b), and therefore, negligible. Symbols in both figures indicate results of direct measurement of $\expct{A([\ps])}_\text{UPO}$ and $pN\sigma(A([\ps]))_\text{UPO}^2$, respectively, by Monte Carlo simulations with $1,000,000$ samples after $100$ steps of transients. Corresponding standard deviations are denoted by error bars. Black dashed curve in (a) represents the exact value $\expct{A}_\mu$. Note that plots for $p=32$ and $64$ are nearly at the same place. Lines and symbols for fixed $p$ and $N$ are within the range of statistical errors, and thus \eqsref{eq:2.9} and \pref{eq:2.10} are confirmed. Furthermore, the two figures show that $\prt{\Psi}{\beta}(1,0)$ and $-\prts{\Psi}{\beta}{2}(1,0)$ converge for sufficiently large $p$, the former of which coincides with the exact value in accordance with \eqref{eq:2.9}, and also for sufficiently large $N$, which indicate that the GMF $\Psi(q,\beta)$ is analytic in the limit $p,N \to \infty$.}
 \label{fig:4}
\end{figure*}%

To calculate numerically the accompanying GMF defined by \eqref{eq:2.7}
 in the limit $p \to \infty$ with fixed $N$, or $N \to \infty$ with fixed $p$,
 it is well known that the zeta function method is
 a powerful tool to accomplish it
 \cite{Cvitanovic_etal-1,Politi_Torcini-1}.
In the present analysis, however,
 we keep both $p$ and $N$ finite in order to maintain
 the formal equivalency between space and time in \eqref{eq:3.5}
 and to exploit knowledge on the equilibrium statistical mechanics.
Since the GMF has the same form as the Helmholtz free energy
 and we know that the GMF at physical situation is zero
 in systems without escape (see \eqref{eq:2.8}),
 we adopt the computational method
 to calculate the difference of the free energy \cite{Frenkel-1}.
From \eqsref{eq:2.6} and \pref{eq:2.7}, we obtain
\begin{equation}
 \e^{-pN \Delta \Psi} = \Expct{ \e^{-pN[\Delta q \lambda([\ps]) + \Delta \beta A([\ps])]}}_{q,\beta},  \label{eq:3.6}
\end{equation}
 where $\Delta \Psi \equiv \Psi(q+\Delta q,\beta+\Delta\beta)-\Psi(q,\beta)$
 and $\expct{\cdots}_{q,\beta}$ denotes the UPO ensemble average
 with the probability distribution
 $f([\ps]) = \e^{-pN[q \lambda([\ps]) + \beta A([\ps])]}$.
The evaluation of the RHS of \eqref{eq:3.6} can be carried out
 by the Monte Carlo method,
 similar to that used in \cite{Kawasaki_Sasa-1}.
The configuration space is a $(1+1)$-dimensional $N \times p$ lattice,
 each direction of which corresponds to space and time, respectively,
 and a spin $s_i^t=\pm 1$ is assigned to each lattice point.
Then the UPO ensemble can be produced by the Metropolis algorithm
 with the ``Hamiltonian'' $pN[q \lambda([\ps]) + \beta A([\ps])]$,
 or the content of the brackets in \eqref{eq:3.5} for the Bernoulli CML,
 by which we can obtain the ensemble average and thus $\Delta \Psi$.
One may think that plausible results are not available
 due to the exponentially increasing number of UPOs with $N$,
 which inevitably restricts the reachable period $p$ to be rather short.
However, it does not affect since accuracy of the Monte Carlo sampling
 is determined by the proportion of dominant orbits in the ensemble average,
 which depends on the product $pN$
 (see \eqref{eq:3.6} and the form of $f([\ps])$ below).
That is to say, the shortness of sampled orbits
 can be compensated by the large number of DOFs.
We can therefore compute the GMF by means of the Monte Carlo method,
 over a wide range of $(q,\beta)$ through the repetition of this step.
Note that the step size $\Delta q$ and $\Delta \beta$ must be
 sufficiently small, otherwise the dominant contribution to the average
 is sustained by the region where $f([\ps])$ is very small and
 hence a sampling during a practicable Monte Carlo run yields
 an inadequate result.
The adequacy of Monte Carlo samplings can be checked
 by plotting $H(\lambda,A)$ by means of \eqref{eq:2.18} and \pref{eq:2.19}
 and seeing that it satisfies the properties of $H(\lambda,A)$,
 such as $0 \leq H(\lambda,A) \leq \lambda$ and the concavity.
We actually confirmed in Fig. \ref{fig:4}
 that the GMF obtained in this way satisfies the relations
 \pref{eq:2.9} and \pref{eq:2.10}.

\subsection{Analysis of the phase transition}  \label{sec:3.c}

\begin{figure}[t]
 \includegraphics[clip]{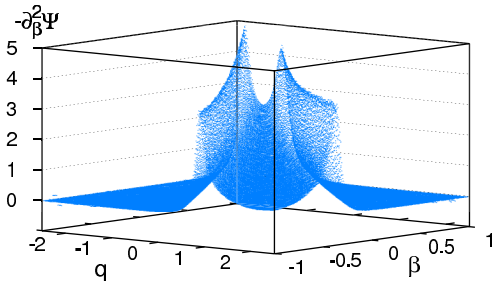}
 \caption{(Color online) The 2nd derivative of the GMF $-\prts{\Psi}{\beta}{2}$ with $k=1, p=N=16$ in the 1D Bernoulli CML. $\Psi(q,\beta)$ is obtained from \eqref{eq:3.6} with $\Delta q=0.02$ and $\Delta \beta = 0.01$. The ensemble average in \eqref{eq:3.6} is performed over $50,000$ Monte Carlo steps after $100$ steps discarded as transients. The derivative $\prts{\Psi}{\beta}{2}$ is yielded by the three-point formula. The figure shown above is smoothed by taking its moving average over $5 \times 5$ data points.}
 \label{fig:5}
\end{figure}%

As is seen from \eqref{eq:3.5},
 the 1D Bernoulli CML involves a 2-dimensional array of spins
 with the short-range interaction in the spatio-temporal configuration space.
It suggests the occurrence of phase transitions
 with finite values of the parameters.
It is indeed the case, which is demonstrated by varying $q$ and $\beta$
 and plotting the 2nd derivative of the ``free energy,''
 $\prts{\Psi}{\beta}{2}$, as is shown in Fig. \ref{fig:5}.
Although we can provide no decisive statement
 about the occurrence of phase transitions
 from finite-size numerical simulations,
 the two sharp peaks in Fig. \ref{fig:5} clearly indicate it,
 which is confirmed by observing
 that they grow more acute as the system size $p$ and $N$ increases.
Therefore the 1D Bernoulli CML is shown to exhibit
 phase transitions in the 2-dimensional space-time.
These transitions, brought about
 by varying the temperature parameters in the thermodynamic formalism,
 are called q-phase transitions
 in the context of dynamical systems with few DOFs \cite{Hata_etal-1}.
Moreover the existence of the q-phase transitions can be analytically shown
 in the weak-interaction limit $k \to 0$.
This can be seen if we neglect $O(k^2)$
 in the argument of the exponential function in \eqref{eq:3.5}, namely
\begin{equation}
 Z_{q,\beta} \simeq \( \frac{1}{2} \)^{pNq/2} \sum_{\{s_{j,k}\}} \exp \[ \( \beta+\frac{kq}{2} \) \sum_\text{n.n.} s_{j,k} s_{j',k'} \],  \label{eq:3.7}
\end{equation}
 which results in the canonical partition function
 of the 2D Ising model,
 where the presence of the 2nd order phase transition
 is certified \cite{Onsager-1}.

Now we mention the meaning of q-phase transitions
 observed in the space-time configuration space
 in terms of the Landau picture of continuous phase transitions.
Here we do not consider the dependence on a macroscopic observable $A(\px)$
 for the sake of simplicity.
The rewriting to the thermodynamic formalism with $A(\px)$ is straightforward.
First we expand $H(\lambda)$ around the temperature $q=q_0$ as follows
\begin{equation}
 H(\lambda) \simeq H_0 + q_0 \lambda - \left[ B(\lambda-\lambda_0)^2 + C(\lambda-\lambda_0)^4 \right],  \label{eq:3.8}
\end{equation}
 where $B \geq 0$ and $C > 0$ because of concavity.
The minimum principle \pref{eq:2.17} yields
 the Lyapunov exponent at the temperature $q$,
 namely $\lambda(q) = \prt{\Psi}{q} \simeq \lambda_0 - (q-q_0)/2B$.
Therefore the 2nd derivative of the ``free energy'' is
 $-\prts{\Psi}{q}{2}=1/2B$, which shows that
 $B$ goes to zero as $q$ approaches a 2nd order q-phase transition point.
Further, the probability to find a positive finite-time Lyapunov exponent
 per 1 DOF $\lambda$ is written as
\begin{equation}
 P_q(\lambda) \propto \e^{-pN[q\lambda-H(\lambda)]} \propto \e^{-pNB[\lambda-\lambda(q)]^2},  \label{eq:3.9}
\end{equation}
 for large $p$.
Thus the occurrence of q-phase transitions involves
 the breakdown of the central limit theorem for
 finite-time Lyapunov exponents and/or finite-time average of
 macroscopic observable $A(\px)$.
It can be understood by the fact that
 the correlation length and time diverge at the 2nd order transition point.
Note that, as the usual Landau theory,
 the above statement cannot be applied at the vicinity of the transition point
 $2pN B^2 \lesssim C$ \cite{Landau_Lifshitz-1}.
Instead, the occurrence of q-phase transitions is ascribed
 to the existence of sharp corners in the function $H(\lambda)$,
 which implies an anomaly in the UPO distribution
 with respect to the Lyapunov exponent $\lambda([\px])$.
Since UPOs form the skeleton of the chaotic invariant set
 \cite{Cvitanovic_etal-1},
 this means a system accompanying q-phase transitions has the invariant set
 with a global anomalous structure.
It is worth remarking that q-phase transitions in chaos with few DOFs
 indicate \textit{local} singularities of the attractor,
 where hyperbolicity is lost \cite{Hata_etal-1},
 whereas q-phase transitions in extended chaos treated here
 signify \textit{global} ones,
 that are non-analyticities in the distribution function of UPOs,
 which arise without losing hyperbolicity.

\begin{figure}[t]
 \includegraphics[clip]{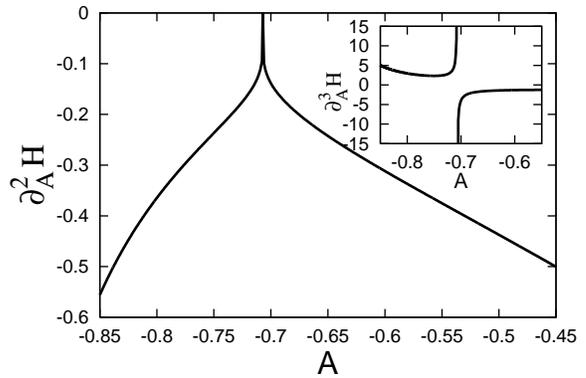}
 \caption{The 2nd and 3rd (inset) derivatives of the ``entropy function'' $H(A)$ in the weak-interaction case $k \ll 1$, which indicate an anomaly in the UPO distribution of the 1D Bernoulli CML. The exact solution of the 2D Ising model \cite{Onsager-1} is used to plot these curves. Note that $H(A)$ does not depend on $k$ if we neglect $O(k^2)$.}
 \label{fig:6}
\end{figure}

The non-analyticity of the UPO distribution in the 1D Bernoulli CML
 can be explicitly confirmed if we consider the case $k \ll 1$,
 in which we can refer to the exact solution
 of the 2D Ising model \cite{Onsager-1} as is seen in \eqref{eq:3.7}.
In this case, a one-to-one correspondence
\begin{equation}
 \lambda([\ps]) \simeq \frac{1}{2}kA([\ps]) + \frac{1}{2} \log 2,  \label{eq:3.a}
\end{equation}
 reduces $H(\lambda,A)$ to a univariate function $H(A)$.
Note that a symbol $\simeq$ here and in \eqsref{eq:3.b} and \pref{eq:3.c} below
 indicates that both sides of the symbol are equal
 as long as we neglect $O(k^2)$.
Let $Z_\text{Ising}$ the partition function of the 2D Ising model per 1 spin
 in the thermodynamic limit, namely
%\begin{multline}
% Z_\text{Ising}(J) = 2 \cosJ (2J) \\ \times \exp \left\{ \frac{1}{2\pi} \int_0^\pi \log \[ \frac{1}{2} \left( 1+ \sqrt{1-k_1(J)^2 \sin^2\varphi} \right) \] \rd\varphi \right\},  \label{eq:3.a}
%\end{multline}
% where $k_1(J) \equiv 2\sinh(2J)/\cosh^2(2J)$.
 $Z_\text{Ising}(J) \equiv \lim_{n \to \infty} \[ \sum_{\{s_{j,k}\}} \exp \( J \sum_\text{n.n.} s_{j,k} s_{j',k'} \) \]^{1/n}$,
 where $n$ is the number of spins.
Then we obtain
 from \eqsref{eq:2.7}, \pref{eq:2.17}, \pref{eq:2.18},
 \pref{eq:3.7}, and \pref{eq:3.a}
 the following relations in the limit $p,N \to \infty$:
\begin{align}
 &A(q,\beta) \simeq -\frac{1}{2} \prt{}{\beta} \log Z_\text{Ising} (\beta+kq/2)  \notag \\
% &\phantom{A(q,\beta)} = -\frac{1}{2} \coth (2\beta + kq) \left[ 1+\frac{2}{\pi} k_1'' K_1 \right]_{J=\beta+kq/2}  \notag \\
 &\phantom{A(q,\beta)} \equiv f_A(\beta+kq/2),  \label{eq:3.b} \\
 &H(A) \simeq \left( \beta + \frac{kq}{2} \right) A + \frac{1}{2} \log Z_\text{Ising} (\beta+kq/2)  \notag \\
 &\phantom{H(A)} = f_A^{-1}(A) A + \frac{1}{2} \log Z_\text{Ising} (f_A^{-1}(A)),  \label{eq:3.c}
\end{align}
% where $k_1'' \equiv 2\tanh^2(2J)-1$
% and $K_1$ is the complete elliptic integral
% $K_1 \equiv K(k_1) \equiv \int_0^{\pi/2} \( 1-k_1^2 \sin^2\varphi \)^{-1/2} \rd\varphi$ with $k_1 \equiv 2 \sinh(2J)/\cosh^2(2J)$.
 where, recalling $A([\ps])$ translates
 into the energy of the 2D Ising model,
 the function $f_A(x)$ is monotonic and thus its inverse is well defined.
The above 2 equations give a simple relation
 $\prts{H}{A}{2} = \( \prt{A}{\beta} \)^{-1}$,
 i.e. the reciprocal specific heat.
Since the specific heat of the 2D Ising model
 logarithmically diverges at criticality,
 the 2nd derivative of $H(A)$ has a sharp corner
 and the 3rd derivative diverges, as are shown in Fig. \ref{fig:6}.
We can of course make the same statement on the function $H(\lambda)$
 because of \eqref{eq:3.a}.
Therefore the anomalous UPO distribution
 actually exists in the 1D Bernoulli CML at least for $k \ll 1$,
 and doubtless for all $k$,
 since q-phase transitions are always numerically observed.

\begin{figure*}
 \includegraphics[clip]{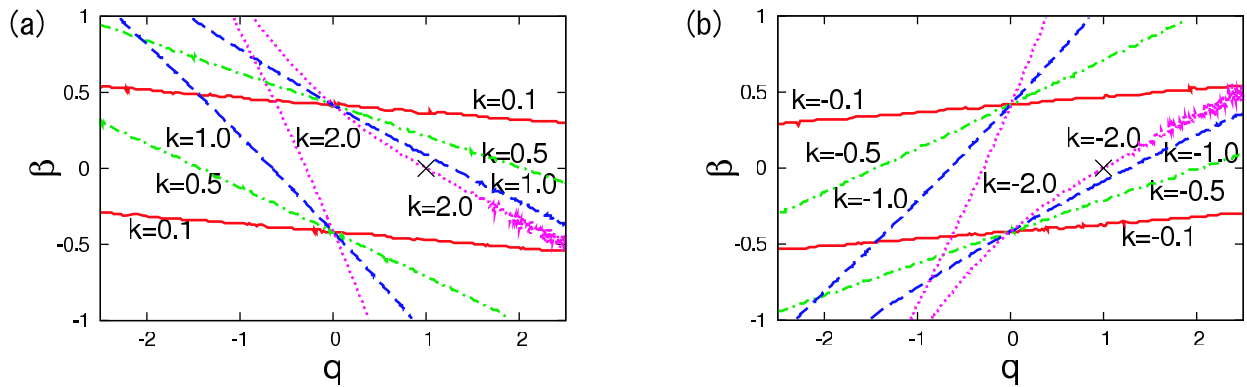}
 \caption{(Color online) q-phase transition curves of the 1D Bernoulli CML with $p=N=16$ and (a) $k=0.1, 0.5, 1.0, 2.0,$ (b) $k=-0.1, -0.5, -1.0, -2.0$, which are indicated by a red solid line, green dot-and-dash line, blue dashed line, and purple dotted line, respectively. The black cross is located on the physical situation $(q,\beta)=(1,0)$. These transition curves are obtained by detecting local maxima of $-\prts{\Psi}{\beta}{2}$ in $q$ and $\beta$ direction, separately, and then eliminating false maxima which come from statistical errors in the Monte Carlo samplings and the finite-size effect. They can be distinguished from rounded-off singularity by examining their continuity and dependence on the system size $p$ and $N$. See also the caption of Fig. \ref{fig:5} for the way to obtain $\prts{\Psi}{\beta}{2}$. The numbers of Monte Carlo samplings are $20,000$, $20,000$, $50,000$, and $100,000$ for $|k|=0.1, 0.5, 1.0, 2.0$ respectively.}
 \label{fig:7}
\end{figure*}%

On the other hand, as mentioned in Sec. \ref{sec:3.a},
 the CML considered here is equivalent to the \textit{1D} Ising model,
 so that \textit{no} Ising phase transition occurs
 at a finite value of the interaction parameter $k$.
It means that, with \eqsref{eq:2.9} and \pref{eq:2.10},
 the GMF $\Psi(q,\beta)$ in the limit $p, N \to \infty$
 is analytic at the physical situation $(q,\beta) = (1,0)$
 and the CML shows no q-phase transition at that point.
Indeed Fig. \ref{fig:7} is a phase diagram for several values of $k$
 and we can see that the transition curves do not go through
 $(q,\beta) = (1,0)$ for not so large $k$.
That is, it is true that the anomalous part in the UPO distribution exists,
 but at finite $k$ those UPOs are hidden as non-dominant terms
 in the partition function $Z_{1,0} = \sum_{[\px]} \e^{-pN\lambda([\px])}$
 and their non-analyticity is overwhelmed by the other,
 analytic and dominant terms.
However, Fig. \ref{fig:7} shows that,
 as $|k|$ is increased and goes to infinity,
 the transition curves move and finally reach the physical situation.
In other words the anomalous UPOs become the dominant terms,
 and at that moment, the Ising phase transition occurs
 and the non-analyticity is uncovered.
This is justified by the fact that $-\prts{\Psi}{\beta}{2}$ specifies
 the lower bound of the fluctuation of the Ising energy, or specific heat,
 so the occurrence of q-phase transition at the physical situation
 just means the actual Ising transition.
Our consideration reveals the role of the anomalous UPO distribution
 as a ``seed'' of the Ising transition, which is ordinarily hidden.
The two transition curves,
 observed at each $k$ in Fig. \ref{fig:7},
 correspond to the ferromagnetic (upper curve)
 and antiferromagnetic (lower curve) transition, respectively,
 which can be understood by comparing transition curves
 for different $k$ in Figs. \ref{fig:7} (a) and (b).

Finally, we add one comment on
 the numerical observation by Kawasaki and Sasa \cite{Kawasaki_Sasa-1}.
In order to explain the reproduction of macroscopic quantities
 in turbulence from a single UPO \cite{1UPO-1},
 they numerically showed that the standard deviation of the Ising energy
 calculated from one UPO, i.e. $\sigma(A([\px]))_\text{UPO}$,
 goes to zero as the system size $N$ increases in the 1D Bernoulli CML.
This is proved by the following facts.
Since the model satisfies the scaling hypothesis \pref{eq:2.14}
 and it does not exhibit a q-phase transition at the physical situation
 for finite values of $k$,
 the GMF $\Psi(q,\beta)$ is assured to be well-defined and analytic
 in the limit $p \to \infty$ and/or $N \to \infty$.
The validity of \eqsref{eq:2.9} and \pref{eq:2.10} in both limits is
 actually suggested by means of Monte Carlo calculations
 as is shown in Fig. \ref{fig:4}.
Therefore, by taking the limit $N \to \infty$ in \eqref{eq:2.10},
 we obtain $\sigma(A([\px]))_\text{UPO} \to 0$ even for a finite period $p$.
This is what Kawasaki and Sasa numerically observed \cite{Kawasaki_Sasa-1},
 and might be a ground for the macroscopic reproduction
 in turbulence \cite{1UPO-1}.
In other words,
 any hyperbolic extended systems which satisfy \eqref{eq:2.14},
 or the large deviation theorem, possess this property.
We can also see from \eqref{eq:2.10} and Fig. \ref{fig:4}(b)
 that the accuracy of a single UPO estimate,
 i.e. standard deviation $\sigma(A([\px]))_\text{UPO}$,
 asymptotically scales as $(pN)^{-1/2}$.
Note that, however, the period must not be too short
 ($p \gtrsim 32$ in the case of the 1D Bernoulli CML)
 in order to regard the UPO average $\expct{A([\px])}_\text{UPO}$
 as a good approximation of the turbulent average
 $\expct{A}_\mu$ (see Fig. \ref{fig:4} and \eqref{eq:2.4}).

\section{Analysis of the 1D repelling CML \protect\\ -solvable case-}
\label{sec:4}

The UPO expansion and the thermodynamic formalism
 dealt with in Sec. \ref{sec:2} are also applicable
 to repelling systems insofar as
 we concentrate our attention into the dynamics on chaotic invariant sets.
A modification is required only on \eqref{eq:2.8},
 which is replaced by
\begin{equation}
 \lim_{p\to\infty} \Psi(1,0) = \alpha,  \label{eq:4.1} \\
% K_q = \frac{1}{q-1} \lim_{p\to\infty} [ \Psi(q,0)-q\alpha ],  \label{eq:4.2}
\end{equation}
 where $\alpha$ is the escape rate per 1 DOF of the repeller,
 i.e. $Z_{1,0} = \sum_{[\px]} \e^{-pN\lambda([\px])} \sim \e^{-pN\alpha}$.
The space-time ``Hamiltonian'' $pN\lambda([\px])$
 can be now constructed at will
 without the strong constraint of \eqref{eq:2.8},
 hence a solvable model is available.

\begin{figure}[t]
%WinTpicVersion3.08
\unitlength 0.1in
\begin{picture}( 30.3000, 28.0500)( -2.2000,-28.6500)
% BOX 2 0 3 0
% 2 500 500 2500 2500
% 
\special{pn 8}%
\special{pa 500 500}%
\special{pa 2500 500}%
\special{pa 2500 2500}%
\special{pa 500 2500}%
\special{pa 500 500}%
\special{fp}%
% VECTOR 2 0 3 0
% 4 250 1500 2750 1500 1500 2750 1500 250
% 
\special{pn 8}%
\special{pa 250 1500}%
\special{pa 2750 1500}%
\special{fp}%
\special{sh 1}%
\special{pa 2750 1500}%
\special{pa 2684 1480}%
\special{pa 2698 1500}%
\special{pa 2684 1520}%
\special{pa 2750 1500}%
\special{fp}%
\special{pa 1500 2750}%
\special{pa 1500 250}%
\special{fp}%
\special{sh 1}%
\special{pa 1500 250}%
\special{pa 1480 318}%
\special{pa 1500 304}%
\special{pa 1520 318}%
\special{pa 1500 250}%
\special{fp}%
% STR 2 0 3 0
% 3 2810 1520 2810 1620 2 0
% $x_i^t$
\put(28.1000,-16.2000){\makebox(0,0)[lb]{$x_i^t$}}%
% STR 2 0 3 0
% 3 1290 130 1290 230 2 0
% $x_i^{t+1}$
\put(12.9000,-2.3000){\makebox(0,0)[lb]{$x_i^{t+1}$}}%
% STR 2 0 3 0
% 3 480 1380 480 1480 3 0
% $-1$
\put(4.8000,-14.8000){\makebox(0,0)[rb]{$-1$}}%
% STR 2 0 3 0
% 3 2540 1370 2540 1470 2 0
% $+1$
\put(25.4000,-14.7000){\makebox(0,0)[lb]{$+1$}}%
% STR 2 0 3 0
% 3 1310 360 1310 460 2 0
% $+1$
\put(13.1000,-4.6000){\makebox(0,0)[lb]{$+1$}}%
% STR 2 0 3 0
% 3 1540 2470 1540 2570 1 0
% $-1$
\put(15.4000,-25.7000){\makebox(0,0)[lt]{$-1$}}%
% STR 2 0 3 0
% 3 950 2600 950 2700 5 0
% $s_i^t=-1$
\put(9.5000,-27.0000){\makebox(0,0){$s_i^t=-1$}}%
% STR 2 0 3 0
% 3 2050 2600 2050 2700 5 0
% $s_i^t=+1$
\put(20.5000,-27.0000){\makebox(0,0){$s_i^t=+1$}}%
% LINE 2 5 3 0
% 2 500 2850 2500 2850
% 
\special{pn 8}%
\special{pa 500 2850}%
\special{pa 2500 2850}%
\special{ip}%
% LINE 2 0 3 0
% 8 600 2500 800 1500 800 1500 1300 500 1700 2500 2200 1500 2400 500 2200 1500
% 
\special{pn 8}%
\special{pa 600 2500}%
\special{pa 800 1500}%
\special{fp}%
\special{pa 800 1500}%
\special{pa 1300 500}%
\special{fp}%
\special{pa 1700 2500}%
\special{pa 2200 1500}%
\special{fp}%
\special{pa 2400 500}%
\special{pa 2200 1500}%
\special{fp}%
% STR 2 0 3 0
% 3 2110 1350 2110 1450 2 0
% $c$
\put(21.1000,-14.5000){\makebox(0,0)[lb]{$c$}}%
% STR 2 0 3 0
% 3 820 1550 820 1650 2 0
% $-c$
\put(8.2000,-16.5000){\makebox(0,0)[lb]{$-c$}}%
% VECTOR 2 0 3 0
% 8 1000 2800 600 2800 1000 2800 1300 2800 2000 2800 1700 2800 2000 2800 2400 2800
% 
\special{pn 8}%
\special{pa 1000 2800}%
\special{pa 600 2800}%
\special{fp}%
\special{sh 1}%
\special{pa 600 2800}%
\special{pa 668 2820}%
\special{pa 654 2800}%
\special{pa 668 2780}%
\special{pa 600 2800}%
\special{fp}%
\special{pa 1000 2800}%
\special{pa 1300 2800}%
\special{fp}%
\special{sh 1}%
\special{pa 1300 2800}%
\special{pa 1234 2780}%
\special{pa 1248 2800}%
\special{pa 1234 2820}%
\special{pa 1300 2800}%
\special{fp}%
\special{pa 2000 2800}%
\special{pa 1700 2800}%
\special{fp}%
\special{sh 1}%
\special{pa 1700 2800}%
\special{pa 1768 2820}%
\special{pa 1754 2800}%
\special{pa 1768 2780}%
\special{pa 1700 2800}%
\special{fp}%
\special{pa 2000 2800}%
\special{pa 2400 2800}%
\special{fp}%
\special{sh 1}%
\special{pa 2400 2800}%
\special{pa 2334 2780}%
\special{pa 2348 2800}%
\special{pa 2334 2820}%
\special{pa 2400 2800}%
\special{fp}%
% STR 2 0 3 0
% 3 750 2040 750 2140 2 0
% $\Gamma_{--,s_{i+1}^t}$
\put(7.5000,-21.4000){\makebox(0,0)[lb]{$\Gamma_{--,s_{i+1}^t}$}}%
% STR 2 0 3 0
% 3 580 890 580 990 2 0
% $\Gamma_{-+,s_{i+1}^t}$
\put(5.8000,-9.9000){\makebox(0,0)[lb]{$\Gamma_{-+,s_{i+1}^t}$}}%
% STR 2 0 3 0
% 3 1930 2140 1930 2240 2 0
% $\Gamma_{+-,s_{i+1}^t}$
\put(19.3000,-22.4000){\makebox(0,0)[lb]{$\Gamma_{+-,s_{i+1}^t}$}}%
% STR 2 0 3 0
% 3 1760 970 1760 1070 2 0
% $\Gamma_{++,s_{i+1}^t}$
\put(17.6000,-10.7000){\makebox(0,0)[lb]{$\Gamma_{++,s_{i+1}^t}$}}%
% STR 2 0 3 0
% 3 950 2850 950 2950 5 0
% $g_{-,s_{i+1}^t}^{-1} (I)$
\put(9.5000,-29.5000){\makebox(0,0){$g_{-,s_{i+1}^t}^{-1} (I)$}}%
% STR 2 0 3 0
% 3 2050 2850 2050 2950 5 0
% $g_{+,s_{i+1}^t}^{-1} (I)$
\put(20.5000,-29.5000){\makebox(0,0){$g_{+,s_{i+1}^t}^{-1} (I)$}}%
\end{picture}%
 \caption{The local map, given by \eqsref{eq:4.3} and \pref{eq:4.4}, for the repelling CML by Just and Schm\"{u}ser. $\Gamma_{\pm\pm,s_{i+1}^t}$ indicates a slope of each piecewise linear part.}
 \label{fig:8}
\end{figure}
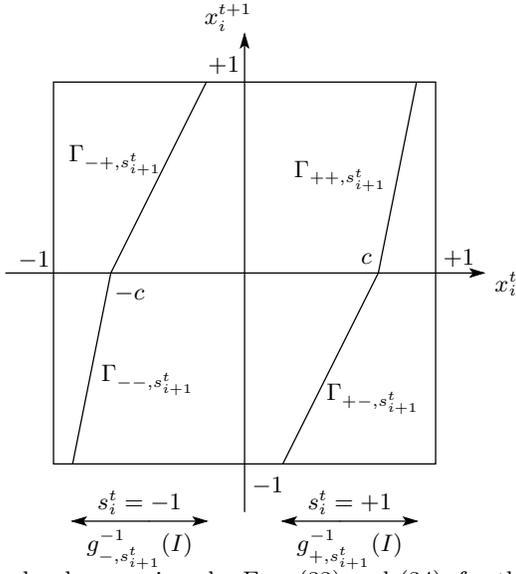

Here we adopt a 1D coupled repeller map lattice introduced
 by Just and Schm\"{u}ser \cite{Just_Schmuser-1}.
Dynamical variables are $x_i^t \in I \equiv [-1,1]$ defined
 at each site $i=0, 1, \cdots, N-1$,
 with a ``spin'' variable
\begin{equation}
 s_i^t \equiv \sgn x_i^t.  \label{eq:4.2}
\end{equation}
The time evolution of $x_i^t$ is yielded by a piecewise linear map
\begin{equation}
 x_i^{t+1} = f(x_i^t;s_{i+1}^t) \equiv \begin{cases} g_{-,s_{i+1}^t}(x_i^t) & \text{if $x_i^t <0$}, \\ g_{+,s_{i+1}^t}(x_i^t) & \text{if $x_i^t>0$}, \end{cases}  \label{eq:4.3}
\end{equation}
 with
\begin{subequations}
\begin{align}
 &g_{-,s}(x) \equiv \begin{cases} \Gamma_{--,s}(x+c) & \text{if $-1<x \leq -c$}, \\ \Gamma_{-+,s}(x+c) & \text{if $-c \leq x < 0$},  \end{cases}  \label{eq:4.4a} \\
 &g_{+,s}(x) \equiv \begin{cases} \Gamma_{+-,s}(x-c) & \text{if $0<x \leq c$}, \\ \Gamma_{++,s}(x-c) & \text{if $c \leq x < 1$},  \end{cases}  \label{eq:4.4b}
\end{align}  \label{eq:4.4}
\end{subequations}
 for some values of slopes $\Gamma_{\pm\pm,s}>1$,
 as functions of the neighboring spin $s$, and a constant $0<c<1$.
The periodic boundary condition $s_N^t=s_0^t$ is considered
 to close the definition.
The local map defined in this way is sketched in Fig. \ref{fig:8}.
The invariant set of this CML can be symbolized again via the partition
\begin{equation}
 V_\ps \equiv g_{s_0,s_1}^{-1} (I) \times g_{s_1,s_2}^{-1}(I) \times \cdots \times g_{s_{N-1},s_0}^{-1}(I),  \label{eq:4.5}
\end{equation}
 where the local partition $g_{s_i,s_{i+1}}^{-1}(I)$ is depicted
 in Fig. \ref{fig:8}.
Hence the first and second symbol in the subscripts of $\Gamma_{\pm\pm,s}$
 indicate a spin of site $i$ at time $t$ and $t+1$ respectively.

\begin{figure}[t]
 \includegraphics[clip]{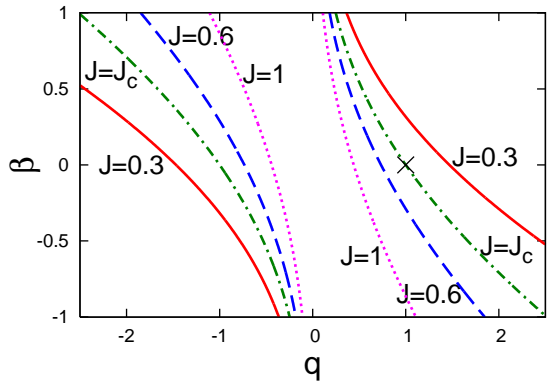}
 \caption{(Color online) q-phase transition curves \pref{eq:4.8} of the 1D repelling CML defined by \eqsref{eq:4.2}-\pref{eq:4.4} and \pref{eq:4.6}, with $J=0.3, J_c \approx 0.44, 0.6, 1.0$, which are drawn with a red solid line, green dot-and-dash line, blue dashed line, and purple dotted line, respectively. The black cross indicates the physical situation $(q,\beta)=(1,0)$. The transition curves for negative $J$ are obtained by reflecting the figure over $q$- or $\beta$-axis.}
 \label{fig:9}
\end{figure}

Now we apply the thermodynamic formalism to the model.
The slopes $\Gamma_{s_i^t s_i^{t+1},s_{i+1}^t}$,
 which prescribe both the local dynamics
 and the interaction between neighboring sites, can be chosen arbitrarily
 provided that the local map does not cross the boundary $x=0,\pm 1$.
Here we choose the simplest form
 after Just and Schm\"{u}ser \cite{Just_Schmuser-1},
\begin{equation}
 \Gamma_{s_i^t s_i^{t+1},s_{i+1}^t} = \exp \[ -J s_i^t \( s_i^{t+1} + s_{i+1}^t \) + e_0 \],  \label{eq:4.6}
\end{equation}
 where $J$ and $e_0$ are some constants.
A macroscopic quantity is set to be the Ising energy again,
 namely \eqref{eq:3.4}.
Thus, the partition function \pref{eq:2.6} for this model
 is calculated as
\begin{widetext}
\begin{equation}
 Z_{q,\beta} = \sum_{[\ps]} \exp \left\{ \sum_{i=0}^{N-1} \sum_{t=0}^{p-1} \[ (qJ+\beta) s_i^t s_{i+1}^t + qJ s_i^t s_i^{t+1} \] \\ -pNqe_0 \right\},  \label{eq:4.7}
\end{equation}
\end{widetext}
 which is nothing but the canonical partition function
 for the 2D Ising model on the square lattice with anisotropic interaction.
Note that the Bernoulli CML treated in the previous section
 results in the 2D Ising model only at the weak interaction limit,
 whereas for the repelling CML it holds for all $q$ and $\beta$.
By setting $(q,\beta)=(1,0)$, i.e. physical situation, in \eqref{eq:4.7},
 the model turns out to show the 2D Ising transition
 at $J=J_c \equiv \frac{1}{2}\ln(1+\sqrt{2})$
 as is shown in \cite{Just_Schmuser-1}.
Moreover, the existence and location
 of the q-phase transition curve $(q_c,\beta_c)$ is also
 given exactly by Onsager's celebrated paper \cite{Onsager-1} as
\begin{equation}
 \beta_c = \frac{1}{2} \Arcsinh\[ 1/\sinh(2q_c J) \] -q_c J.  \label{eq:4.8}
\end{equation}
Figure \ref{fig:9} shows its phase diagram for several values
 of the coupling constant $J$.
In this case, the transition curve passes the physical situation
 $(q,\beta)=(1,0)$ linearly as $J$ goes through $J_c$,
 which can be explicitly written from \eqref{eq:4.8} as
\begin{subequations}
\begin{align}
 &\beta_c \simeq -2(J-J_c),  &&\text{for fixed $q_c$,}  \label{eq:4.9a} \\
 &q_c \simeq 1-(J-J_c)/J_c,  &&\text{for fixed $\beta_c$.}  \label{eq:4.9b}
\end{align}  \label{eq:4.9}
\end{subequations}
This linear dependence of the q-phase transition point $(q_c,\beta_c)$
 on the control parameter $J$
 at the vicinity of the actual phase transition point $J=J_c$
 results from the fact that the transition observed here
 is not a marginal one.
Hence this relation between the two transition points
 is expected to be general.

On the anomalous structure of the invariant set
 with respect to the UPO distribution
 and its role in the occurrence of the phase transition,
 the same statement as the previous section holds,
 which can be demonstrated directly for this model
 since the rigorous solution is available.

\section{Discussion}  \label{sec:5}

The intimate relation between q-phase transitions and
 phase transitions in the sense of statistical mechanics is investigated
 on the basis of the thermodynamic formalism and the UPO expansion.
Since mathematically
 the partition function \pref{eq:2.6} has the identical form
 to that of the canonical statistical mechanics with $q$ and $\beta$
 as inverse temperature,
 many useful relations in equilibrium physics,
 such as \eqsref{eq:2.9}, \pref{eq:2.10}, and \pref{eq:2.18},
 also hold in extended chaotic systems,
 which can be far from equilibrium.
Although similar relations have been already pointed out
 for dynamical systems with few DOFs by several authors
 \cite{Fujisaka_Inoue-1,Cvitanovic_etal-1,Hata_etal-1},
 we reconstructed it for extended systems
 with concerning the number of DOFs $N$ explicitly.
By that means the analogy is kept with the equilibrium statistical mechanics
 of several-dimensional systems.
A richer harvest may be reaped from it,
 as long as attention is paid to the strict constraint of \eqref{eq:2.8}
 for systems without escape.

As regards phase transitions,
 anomalously distributed UPOs turn out to be responsible,
 which show sharp corners in the distribution function and 
 can be visualized in terms of q-phase transition.
The anomalous part exists in systems with transitions,
 over the range of control parameters
 where the topological structure of an attractor does not change.
It is ordinarily hidden as non-dominant terms
 in the partition function $Z_{1,0}$
 and no critical behavior is observed there.
The actual transition occurs when the control parameters are varied,
 the UPO distribution is changed
 and finally the anomalous UPOs become a dominant part.
In this sense we call the anomalous part of the UPO distribution
 ``seed'' of phase transitions.

One question may arise here:
 ``What brings this anomalous UPO distribution
 to dynamical systems with phase transitions?''
The answer is clear for the two Ising-like systems considered in this paper,
 where the origin of phase transitions is by construction well known
 from the knowledge of equilibrium statistical mechanics:
 the competition between interaction energy and entropy is relevant.
%Let us recall that the free energy calculated under some fixed energy $E$,
% defined by $\e^{-\beta F(E)} \equiv \sum' \e^{-\beta E}$
% where the sum is restricted over microstates of energy $E$,
% can be written as $F(E) = E-TS(E)$.
%For strongly ferromagnetic configurations, the internal energy is low
% while the number of such configurations is very restricted.
%That is, the entropy $S(E)$ is very low
% and the free energy $F(E)$is rather high.
Taking ferromagnet for example,
 the free energy $F(E)=E-TS(E)$ calculated under some fixed energy
 is increased by low entropy
 for strongly ferromagnetic configurations
 (corresponding to low $E$),
 while it is raised by high internal energy
 for strongly paramagnetic configurations (corresponding to high $E$).
It means there are intermediate configurations
 where the two mechanisms compete.
In fact, this competition occurs at one point,
 i.e. at some specific value of $E$, in the thermodynamic limit,
 which brings a sharp corner to the functional form of $S(E)$.
Then the phase transition occurs at a temperature
 which minimizes the free energy at that point.

The role of UPOs in \textit{q-}phase transitions
 --- not in \textit{actual} phase transitions --- is exactly the same
 as that of microstates which we have seen above.
That is, the anomalous distribution of UPOs results
 from the competition between average positive Lyapunov exponent
 and topological entropy.
This mechanism may be widespread
 even among ``natural'' extended chaotic systems,
 because it is reasonable to expect that
 the number of UPOs with plenty of large Lyapunov exponents is very small,
 and that it grows in the same manner as equilibrium microstates
 (see \eqref{eq:2.14} and Ref. \cite{Beck-1,Kubo-1}).
Note that the existence of symbolic dynamics is also \textit{not} required,
 since the underlying basis described in Sec. \ref{sec:2}
 is constructed generally for hyperbolic maps.

Taking into account the above considerations
 and the aforesaid similarities
 in statistics of macroscopic observables
 such as \eqsref{eq:2.9} and \pref{eq:2.10},
 we can propose a definition of phase transitions
 and their order in extended dynamical systems in the Ehrenfest's sense:
 phase transitions are associated with the singularity of the GMF
 at the physical situation $(q,\beta)=(1,0)$.
The transition can be said to be of $n$-th order
 if an $n$-th derivative of the GMF with respect to $q$ or $\beta$
 does not exist or has a discontinuity,
 and hence the system is accompanied by $n$-th order q-phase transition.
This is a mathematically simple-minded statement
 as well as that based on the non-uniqueness of a natural measure.
Moreover it is worth remarking that the proposed definition
 can clearly characterize both first and higher order transitions,
 while definitions in the Gibbsian sense have some ambiguity
 when it comes to treat higher order transitions.
Though the new definition has also several problems
 which will be mentioned later,
 it can be used to classify phase transitions out of equilibrium
 and to investigate their nature further.
This is the main proposition of this paper.

The second outcome is on the propriety of treating
 locally defined control parameters as macroscopic ``temperature.''
As we have seen in Sec. \ref{sec:3} and \ref{sec:4},
 systems which exhibit a phase transition have an anomalous UPO distribution,
 the position of whose non-analytic part is specified
 by critical nominal temperatures $q_c$ and $\beta_c$.
Therefore, our observation that they vary with local control parameters
 means that a change in local parameters leads to a change
 in ``macroscopic temperature''
 through the non-analyticity of the UPO distribution.
This macroscopic temperature actually takes part in phase transitions,
 since it crosses through the physical situation $(q,\beta)=(1,0)$
 when a transition occurs,
 and since it mathematically works in the same way as real temperature
 in equilibrium systems
 (cf. \eqsref{eq:2.6}, \pref{eq:2.9}, and \pref{eq:2.10}).

Let us then discuss the replacement of ``temperature''
 by local control parameters around transition points.
Let $P$ denote some control parameter
 in an extended dynamical system with a phase transition.
As we have already seen in Sec. \ref{sec:4},
 the relation between parameter $P$
 and q-phase transition point $(q_c,\beta_c)$
 at the vicinity of the transition point $P=P_c$
 is expected to be linearly dependent
\begin{subequations}
\begin{align}
 &\beta-\beta_c \simeq C_1(P-P_c),  &&\text{for fixed $q_c$,}  \label{eq:5.1a} \\
 &q-q_c \simeq C_2(P-P_c),  &&\text{for fixed $\beta_c$,}  \label{eq:5.1b}
\end{align}  \label{eq:5.1}
\end{subequations}
 for transitions which are not marginal.
Here we set $q$ and $\beta$ at the physical situation $(q,\beta)=(1,0)$.
Therefore, as far as some universal relation in equilibrium physics
 is concerned which is not affected by microscopic details of models,
 e.g. critical behavior,
 the same relation may hold in extended chaotic systems
 by replacing inverse temperature $(1/k_B T-1/k_B T_c)$ with $C_1(P-P_c)$.
% or temperature $(T-T_c)$ with $-k_B T_c^2 C_1 (P-P_c)$.
A similar statement could be also said
 on the positive Lyapunov exponent per 1 DOF,
 in which case \eqref{eq:5.1b} is used for the replacement.
Conditions which should be satisfied at least by the relation are that
 (1) it is about some macroscopic quantities
 obtained by differentiating a free energy,
 and that (2) its mathematical expression itself is insensitive to variations
 in the control parameter $P$.
The difference in the relation between 2nd or higher order moments
 and derivatives of the ``free energy'' from its counterpart
 in equilibrium statistical mechanics might also have an influence
 --- in extended chaos the derivatives can only tell the lower bound
 of the corresponding moments due to temporal correlation,
 as is seen in \eqref{eq:2.10} ---.
Provided that those restrictions are taken into consideration,
 we believe that the mentioned replacement can be applied to
 a wide range of extended systems.
Note that universal scaling relations
 in critical behavior satisfy the above conditions
 and thus corresponding critical exponents are likely to be kept invariant
 under the replacement of temperature $T$ by the local control parameter $P$.
It can be a basis on which scaling relations indeed work
 under such a replacement in some high-dimensional chaotic systems
 (e.g. \cite{Marcq_Chate_Manneville-2,Marcq_Chate_Manneville-1}).

In order to justify the above arguments on a rigid basis,
 several problems need to be clarified.
To begin with,
 it is unclear in extended chaotic systems
 how common the existence of the anomalous UPO distribution is,
 and also how prevalent its relation to phase transitions is.
The latter is especially important,
 since it involves a change in the UPO distribution
 and thus there is no counterpart in equilibrium statistical mechanics.
Further studies are crucial.

From a fundamental point of view,
 we do not mathematically care in this paper either the existence of
 the two limits $p \to \infty$ and $N \to \infty$,
 their order or the fact that they do not commute.
They are undoubtedly important in order to argue
 spatio-temporal chaos on the mathematically proper basis
 \cite{MathReviews-1,Gielis_MacKay-1,Blank_Bunimovich-1,Bardet_Keller-1,Bricmont_Kupiainen-1}.
An examination of the behavior of infinite-size systems requires 
 that we first take the limit $N \to\infty$ and then $p \to\infty$,
 at variance with usual statistical mechanics
 where the limit is taken over sizes of all dimensions simultaneously.
This may be the reason why some extended chaotic systems
 defined in $d$-dimensional space show critical behaviors
 of $d$-dimensional universality classes
 despite the corresponding $(d+1)$-dimensional configuration space
 \cite{Marcq_Chate_Manneville-1}.
The problem of the incommutability should be considered seriously,
 since the definition of phase transitions
 by means of the singularity of the GMF
 involves both the limit $p,N \to \infty$.

Another problem is on the arbitrariness
 for the choice of a macroscopic quantity $A(\px)$
 when we deal with q-phase transitions with respect to $\beta$.
There is no clear criterion for it,
 except for that $A(\px)$ must be affected by phase transitions :
 its expectation value or fluctuation must show a discontinuity or divergence.
An order parameter of the considered transition is a candidate.
In our examples in Sec. \ref{sec:3} and \ref{sec:4}, however,
 we adopted the quantity which can be regarded as energy
 on the analogy with equilibrium spin systems.
The relation between the choice for $A(\px)$ and
 the behavior of q-phase transition curves near critical points
 remains to be clarified,
 especially about their linear dependence on control parameters
 such as \eqref{eq:5.1}.

In conclusion, the old concept of the thermodynamic formalism
 and the periodic orbit expansion turns out to be useful
 to characterize phase transitions in extended dynamical systems.
Theoretically, one possible definition of phase transitions is proposed,
 which is complementary to the usual definition
 in terms of a change in the number of natural measures.
It can be used to classify and examine non-equilibrium transitions
 in chaotic systems, especially higher order transitions,
 with the help of a suitable technique
 to generate or approximate the UPO ensemble.
Recently developed method such as Ref. \cite{Lan_Cvitanovic-1,Sasa_Hayashi-1}
 might be applied for this purpose.
With regard to experiments and numerical simulations,
 a ground is obtained on which
 we can sometimes treat an externally imposed control parameter
 as macroscopic ``temperature'' around phase transition points.
Although some significant problems are left for future studies,
 this assertion is expected to support discussions
 on universality classes in non-equilibrium systems
 from real and numerical experiments.

\begin{acknowledgements}
The authors gratefully acknowledge sincere comments from H. Tasaki and S. Sasa,
 and beneficial discussions with K. Kaneko, M. Kawasaki, and S. Tatsumi.
One of us (K. T.) would also like to thank K. Nakajima
 for letting him use the PC cluster Cenju for this work.
\end{acknowledgements}


\begin{thebibliography}{99}
%\bibitem{Ott-1}
% E. Ott, \textit{Chaos in Dynamical Systems}, 2nd ed. (Cambridge University Press, 2002).
\bibitem{Ruelle-1}
 D. Ruelle, \textit{Thermodynamic formalism} (Addison-Wesley, 1978).
\bibitem{Beck-1}
 C. Beck and F. Schl\"{o}gl, \textit{Thermodynamics of chaotic systems: an introduction} (Cambridge University Press, 1993).
\bibitem{Halsey_etal-1}
 T. C. Halsey, M. H. Jensen, L. P. Kadanoff, I. Procaccia, and B. I. Shraiman, Phys. Rev. A \textbf{33}, 1141 (1986).
\bibitem{Fujisaka_Inoue-1}
 H. Fujisaka and M. Inoue, Prog. Theor. Phys. \textbf{77}, 1334 (1987); Phys. Rev. A \textbf{39}, 1376 (1989); Phys. Rev. A \textbf{41}, 5302 (1990).
\bibitem{Cvitanovic_etal-1}
 P. Cvitanovi\'{c}, Phys. Rev. Lett. \textbf{61}, 2729 (1988); P.Cvitanovi\'{c} \textit{et al.}, \textit{Chaos : Classical and Quantum} (Niels Bohr Institute, Copenhagen, 2005), available at \verb|http://chaosbook.org/|.
\bibitem{Hata_etal-1}
 H. Hata, T. Horita, H. Mori, T. Morita, and K. Tomita, Prog. Theor. Phys. \textbf{80}, 809 (1988); Prog. Thoer. Phys. \textbf{81}, 11 (1989); T. Horita, H. Hata, H. Mori, T. Morita, K. Tomita, S. Kuroki, and H. Okamoto, Prog. Theor. Phys. \textbf{80}, 793 (1988); S. Sato and K. Honda, Phys. Rev. A \textbf{42}, 3233 (1990).
\bibitem{MathReviews-1}
  See e.g. L. A. Bunimovich and Ya. G. Sinai, Nonlinearity \textbf{1}, 491 (1988); for hyperbolic maps, e.g. M. Jiang, Nonlinearity \textbf{8}, 631 (1995); for reviews, e.g. L. A. Bunimovich, Physica D \textbf{103}, 1 (1997); J. Bricmont and A. Kupiainen, Physica D \textbf{103}, 18 (1997).
\bibitem{NumericWorks-1}
 See, e.g. P. Manneville, in \textit{Macroscopic Modelling of Turbulent Flows}, Lecture Notes in Physics Vol. 230, edited by U. Frisch, J. B. Keller, G. Papanicolaou, and O. Pironneau (Springer-Verlag, Berlin, 1985), p. 319; F. Christiansen, P. Cvitanovi\'{c}, and V. Putkaradze, Nonlinearity \textbf{10}, 55 (1997).
\bibitem{Politi_Torcini-1}
 A. Politi and A. Torcini, Phys. Rev. Lett. \textbf{69}, 3421 (1992).
\bibitem{1UPO-1}
 G. Kawahara and S. Kida, J. Fluid Mech. \textbf{449}, 291 (2001); S. Kato and M. Yamada, Phys. Rev. E \textbf{68}, 025302(R) (2003); L. van Veen, S. Kida, and G. Kawahara, Fluid Dyn. Res. \textbf{38}, 19 (2006).
\bibitem{Kawasaki_Sasa-1}
 M. Kawasaki and S. Sasa, Phys. Rev. E \textbf{72}, 037202 (2005).
\bibitem{Chate_Manneville-1}
 H. Chat\'{e} and P. Manneville, Europhys. Lett., \textbf{17}, 291 (1992); Prog. Theor. Phys. \textbf{87}, 1 (1992); H. Chat\'{e}, A. Lema\^{i}tre, P. Marcq, and P. Manneville, Physica A \textbf{224}, 447 (1996).
\bibitem{Marcq_Chate_Manneville-2}
 P. Marcq, H. Chat\'{e}, and P. Manneville, Prog. Theor. Phys. Suppl., \textbf{161}, 244 (2006).
\bibitem{Miller_Huse-1}
 J. Miller and D. A. Huse, Phys. Rev. E \textbf{48}, 2528 (1993).
\bibitem{Marcq_Chate_Manneville-1}
 P. Marcq, H. Chat\'{e}, and P. Manneville, Phys. Rev. Lett. \textbf{77}, 4003 (1996); Phys. Rev. E \textbf{55}, 2606 (1997).
\bibitem{Gielis_MacKay-1}
 G. Gielis and R. S. MacKay, Nonlinearity \textbf{13}, 867 (2000); R. S. MacKay, in \textit{Dynamics of coupled map lattices and related spatially extended systems}, Lecture Notes in Physics Vol. 671, edited by J. -R. Chazottes and B. Fernandez (Springer-Verlag, Berlin, 2005), p. 65.
\bibitem{Just_Schmuser-1}
 W. Just, J. Stat. Phys. \textbf{105}, 133 (2001); W. Just and F. Schm\"{u}ser, in \textit{Dynamics of coupled map lattices and related spatially extended systems}, Lecture Notes in Physics Vol. 671, edited by J. -R. Chazottes and B. Fernandez (Springer-Verlag, Berlin, 2005), p. 33.
\bibitem{Blank_Bunimovich-1}
 M. Blank and L. A. Bunimovich, Nonlinearity \textbf{16}, 387 (2003).
\bibitem{Bardet_Keller-1}
 J. -B. Bardet and G. Keller, Nonlinearity \textbf{19}, 2193 (2006).
\bibitem{vanEnter_etal-1}
 A. C. D. van Enter, R. Fern\'{a}ndez, and A. D. Sokal, J. Stat. Phys. \textbf{72}, 879 (1993). See Sec. 2.6.5 and references therein.
\bibitem{Bricmont_Kupiainen-1}
 J. Bricmont and A. Kupiainen, Commun. Math. Phys. \textbf{178}, 703 (1996).
\bibitem{Lebowitz_etal-1}
 J. L. Lebowitz, C. Maes, and E. R. Speer, J. Stat. Phys. \textbf{59}, 117 (1990).
\bibitem{EffectiveTemperature-1}
 P. C. Hohenberg and B. I. Shraiman, Physica D \textbf{37}, 109 (1989); M. S. Bourzutschky and M. C. Cross, Chaos \textbf{2}, 173 (1992); F. Sastre, I. Dornic, and H. Chat\'{e}, Phys. Rev. Lett. \textbf{91}, 267205 (2003).
\bibitem{UPO_Expansion-1}
 T.Kai and K. Tomita, Prog. Theor. Phys. \textbf{64}, 1532 (1980); T. Morita, H. Hata, H. Mori, T. Horita, and K. Tomita, Prog. Theor. Phys. \textbf{79}, 296 (1988); C. Grebogi, E. Ott, and J. A. Yorke, Phys. Rev. A \textbf{37}, 1711 (1988).
\bibitem{Kawasaki-PC}
 M. Kawasaki (private communication).
\bibitem{Lyap.Part.Func.-1}
 M. Sano, S. Sato, and Y. Sawada, Prog. Theor. Phys. \textbf{76}, 945 (1986); J. -P. Eckmann and I. Procaccia, Phys. Rev. A \textbf{34}, 659 (1986).
\bibitem{Sakaguchi-1}
 H. Sakaguchi, Prog. Theor. Phys. \textbf{80}, 7 (1988).
\bibitem{Frenkel-1}
 D. Frenkel, in \textit{Molecular-Dynamics Simulation of Statistical-Mechanical Systems}, Proceedings of the International School of Physics ``Enrico Fermi'', Course 97, edited by G. Ciccotti and W. G. Hoover (North-Holland, Amsterdam, 1986), p. 151.
\bibitem{Onsager-1}
 L. Onsager, Phys. Rev. \textbf{65}, 117 (1944).
\bibitem{Landau_Lifshitz-1}
 L. D. Landau and E. M. Lifshitz, \textit{Statistical Physics}, 3rd ed. (Pergamon Press, Oxford, 1980).
\bibitem{Kubo-1}
 R. Kubo, \textit{Statistical Mechanics: an advanced course with problems and solutions} (North-Holland, Amsterdam, 1965).
\bibitem{Lan_Cvitanovic-1}
 Y. Lan and P. Cvitanovi\'{c}, Phys. Rev. E \textbf{69}, 016217 (2004).
\bibitem{Sasa_Hayashi-1}
 S. Sasa and K. Hayashi, Europhys. Lett. \textbf{74}, 156 (2006).
\end{thebibliography}
\end{document}